\documentclass[10pt,twocolumn,twoside]{IEEEtran}
\renewcommand{\baselinestretch}{1}

\usepackage{color}
\usepackage{theorem}
\usepackage{times,amsmath,epsfig}
\usepackage{amssymb}
\usepackage{subfigure}
\usepackage{cite}
\usepackage{cases}

\usepackage{algorithmic}
\usepackage{algorithm}
\usepackage{needspace}

\usepackage{tikz}
\usetikzlibrary{shapes,arrows}
\input{mysymbol.sty}
\tikzstyle{phantom vertex} = [ ellipse, 
                               anchor = center, 
                               minimum height = 1*\unit, 
                               minimum width  = 1*\unit,
                               inner sep=0pt,
                               anchor=center]
\tikzstyle{red vertex}   = [black, fill = red!20,   phantom vertex, draw]
\tikzstyle{black vertex} = [black, fill = black!20, phantom vertex, draw]
\tikzstyle{blue vertex}  = [black, fill = blue!20,  phantom vertex, draw]
\tikzstyle{green vertex} = [black, fill = green!20,  phantom vertex, draw]
\tikzstyle{yellow vertex} = [black, fill = yellow!20,  phantom vertex, draw]
\tikzstyle{cyan vertex} = [black, fill = cyan!20,  phantom vertex, draw]
\tikzstyle{vertex}       = [draw, phantom vertex]

\tikzstyle{point} = [ellipse, inner sep=0pt, draw, fill=white, anchor = center,
                     minimum height = 0.05*\unit, minimum width  = 0.05*\unit]

\newcommand{\mymod}{\mathrm{mod}_N}

\newtheorem{mytheorem}{\bf Theorem}

\newtheorem{mycorollary}{\bf Corollary}
\newtheorem{mylemma}{\bf Lemma}
\newtheorem{myproposition}{\bf Proposition}
\newtheorem{remark}{\bf Remark}

\title{Blind Identification of Graph Filters}

\author{\IEEEauthorblockN{Santiago Segarra, Gonzalo Mateos, Antonio G. Marques, and Alejandro Ribeiro}
\thanks{Work in this paper is supported by USA NSF CCF-1217963 and Spanish MINECO grant No TEC2013-
	41604-R. S. Segarra and A. Ribeiro are with the Dept. of Electrical and Systems Eng., Univ. of Pennsylvania., G. Mateos is with the Dept. of Electrical and Computer Eng., Univ. of Rochester, A. G. Marques is with the Dept. of Signal Theory and Comms., King Juan Carlos Univ.  Emails: ssegarra@seas.upenn.edu,  gmateosb@ece.rochester.edu, antonio.garcia.marques@urjc.es, and aribeiro@seas.upenn.edu. Part of the results in this paper were presented at the \textit{2015 CAMSAP Workshop}~\cite{SSGMAMAR_camsap15}, and at the \textit{2016 ICASSP Conference}~\cite{SSAMGMAR_icassp16}.}}

\begin{document}
\maketitle

\begin{abstract}%
Network processes are often represented as signals defined on the vertices of a graph. To untangle the
latent structure of such signals, one can view them as outputs of \emph{linear graph filters} modeling
underlying network dynamics. This paper deals with the problem of joint identification of a graph filter and its input signal, 
thus broadening the scope of classical blind deconvolution of temporal and spatial signals to the less-structured graph domain. Given a graph
signal $\bby$ modeled as the output of a graph filter, the goal is to recover the vector of filter coefficients $\bbh$,
and the input signal $\bbx$ which is assumed to be sparse. While $\bby$ is a bilinear function of
$\bbx$ and $\bbh$, the filtered graph signal is also a linear combination of the entries of the lifted rank-one, 
row-sparse matrix $\bbx\bbh^T$. The blind graph-filter
identification problem can thus be tackled via rank and sparsity minimization subject to linear constraints, an 
inverse problem amenable to convex relaxations offering provable recovery guarantees under simplifying assumptions. 
Numerical tests using both synthetic and real-world networks illustrate the merits of the proposed algorithms,
as well as the benefits of leveraging multiple signals to aid the blind identification task.


\end{abstract}


%
\section{Introduction}\label{S:Introduction}

Coping with the challenges found at the intersection of Network Science and Big Data necessitates
broadening the scope beyond classical temporal signal analysis and processing, to also accommodate \emph{signals defined on graphs}~\cite{EmergingFieldGSP,SandryMouraSPG_TSP13,kolaczyk2009book}.
Under the assumption that the signal properties are related to the topology of the graph where they are supported, the goal of graph signal processing (GSP) is to develop algorithms that fruitfully leverage this relational structure, and can make inferences about these relationships when they are
only partially observed~\cite{kolaczyk2009book}.
A suitable way to accomplish these objectives is to rely on the so-called graph-shift operator, which is a matrix 
that reflects the local connectivity of the graph~\cite{SandryMouraSPG_TSP13}.

We consider here that each node has a certain value, and these values are collected across nodes to form a graph signal.
With this definition, graph filters -- which are a generalization of classical time-invariant systems -- are a specific class of operators whose input and output are graph signals (cf. Section \ref{S:Modeling}). Mathematically, graph filters are linear transformations that can be expressed as polynomials of the graph-shift operator~\cite{SandryMouraSPG_TSP14Freq}. 
The polynomial coefficients determine completely the transformation and are referred to as \emph{filter coefficients}. 
Such linear transformations can be implemented via local interactions among nodes, and may be used to model 
e.g., diffusion or percolation dynamics in the network~\cite{ssamar_distfilters_allerton15, segarra2015graphfilteringTSP15, meigraphstructure}. 

\medskip\noindent\textbf{Contributions.} This paper investigates the problem of blind identification of graph filters. Specifically, we are given a graph signal $\bby$ which is assumed to be the output of a graph filter, and seek to \emph{jointly identify} the \emph{filter} coefficients $\bbh$ and the \emph{input signal} $\bbx$ that gave rise to $\bby$.
This is the extension to graphs of the classical problem of blind system identification or blind deconvolution of signals in the time or spatial domains~\cite{ahmed2014BlindDeconvConvex}. Since the inverse problem is ill-posed, we assume that the length of $\bbh$ is small and that $\bbx$ is sparse. This is the case when, e.g., a few seeding nodes inject a signal that is diffused throughout a network~\cite{segarra2015reconstruction, EUSIPCO_our_interp_2015}.
While $\bby$ is a bilinear function of $\bbx$ and $\bbh$, we show that the filtered graph signal is also a linear combination of the entries of the lifted rank-one, row-sparse matrix $\bbx\bbh^T$~\cite{ahmed2014BlindDeconvConvex,LingBiConvexCS}. The blind graph-filter
identification problem can thus be tackled via joint rank and sparsity minimization subject to linear constraints, an approach amenable to convex relaxation~\cite{fazel2001RankMinimization,oymak2015SimultStruct}. Several alternatives are proposed to approach such a relaxation, including generalizations facilitating blind graph-filter identification when multiple outputs (each corresponding to a different input) are available; see also~\cite{LiIdentifiabilityBilinear} for identifiability claims in a setting unrelated to graphs. Under simplifying assumptions, probabilistic recovery conditions are also derived. Together with the proof, effort is devoted towards building intuition on the obtained performance guarantees by identifying graph-related parameters that have a major impact on blind identification, as well as by distilling the fundamental differences relative to the time domain~\cite{LingBiConvexCS}. Numerical tests not only showcase the effectiveness of the proposed algorithm on synthetic and real graphs, but also illustrate that recovery is in practice possible under conditions far less restrictive than those stemming from the analysis in Section \ref{S:Recovery}. 

\medskip\noindent\textbf{Envisioned applications.} The dynamics of opinion formation in social networks can be modeled using distributed linear operations 
implemented by multi-agent systems; see e.g.,~\cite{opinionformation,Hegselmann.opinion}.
Interestingly, graph filters have been used to implement distributedly related linear transformations such
as fast consensus~\cite{SandryMouraICASSP14consensus}, and projections onto the low-rank space of graph-bandlimited signals~\cite{SafaviSPLconensus}. This motivates
adopting the algorithms proposed here to identify those influential actors (i.e., the non-zero entries in $\bbx$) that 
instilled the observed status-quo. Another example of interest is given by structural and functional brain networks, which are becoming increasingly central to the analysis of brain signals. Nodes correspond to regions of interest (ROIs) and their associated information (e.g., their level of neural activity) can be represented as a graph signal. Suppose that an observed brain signal corresponds to the linear combination of a diffused pattern of an originally sparse brain signal (i.e., generated by a few active ROIs). Blind identification amounts then to jointly estimating the desired original brain signal and the combination coefficients. 
In the analysis of epileptic seizure data for instance,
estimating the (sparse) input state can help to identify the ROIs from
where the seizure emanated, which may serve to guide surgical intervention~\cite{epilepsy}. While linear models of processes in the brain are admittedly simplistic, they can still offer informative insights~\cite{linearbrainmodel}. In the same vein, we envision applications in marketing where
e.g.,  social media advertisers want to identify a small set of initiators
so that an online campaign can go viral; in
healthcare policy implementing network analytics to infer hidden needle-sharing networks of 
injecting drug users~\cite{kolaczyk2009book}; or, in environmental monitoring using wireless sensor networks to localize heat or seismic sources~\cite{DLMS}.

\medskip\noindent\textbf{Relation to prior work.} Our ideas are inspired by the work in~\cite{ahmed2014BlindDeconvConvex}, where matrix lifting is used for blind deconvolution of temporal and spatial signals. In the current paper, the linear operator mapping $\bbx\bbh^T$ to the output signal $\bby$ depends on the spectral properties of the graph-shift operator~\cite{SSGMAMAR_camsap15}, a departure from the random (Gaussian or partial Fourier) operators arising with the biconvex compressed sensing approach in~\cite{LingBiConvexCS}. Despite its practical interest, the setup where multiple output signals are observed (each one corresponding to a different sparse input) has received little attention in recent convex relaxation approaches to blind deconvolution~\cite{LiIdentifiabilityBilinear}.

\medskip\noindent\textbf{Paper outline and notation.} Section \ref{S:Modeling} introduces notation and explains how graph signals and filters can be used to model linear diffusion processes. Section \ref{S:BlindIdent} formulates the problem of blind graph-filter identification and proposes several efficient convex relaxations. In particular, Section \ref{S:MultipleOutputs} discusses algorithms for the setup where multiple observed outputs are available. Section \ref{S:Recovery} provides analytical results on the recovery performance, along with graph-specific parameters that affect the recovery guarantees. Numerical experiments illustrating the merits of our approach are presented in Section~\ref{S:Simulations} and concluding remarks are given in Section~\ref{S:Conclusions}.

\medskip\noindent\emph{Notation:} Entries of a matrix $\bbX$ and a (column) vector $\bbx$ are denoted as $X_{ij}$ and $x_i$; but when contributing to avoid confusion, $[\bbX]_{ij}$ and $[\bbx]_i$ are used instead. Operators $(\cdot)^T$, $(\cdot)^H$, $\E{\cdot}$, $\circ$, $\otimes$ and $\odot$ stand for matrix transpose, conjugate transpose (Hermitian), expectation, Hadamard (entry-wise), Kronecker, and Khatri-Rao (column-wise Kronecker) products, respectively. Matrix $\diag(\mathbf{x})$ is diagonal with $[\diag(\mathbf{x})]_{ii}=[\mathbf{x}]_i$;  and $|\cdot|$ is used for the cardinality of a set, and the magnitude of a scalar. The complex conjugate of $x$ is denoted as $\bar{x}$. The $n\times n$ identity matrix is represented by $\bbI_n$, while $\mathbf{0}_n$ stands for the $n\times 1$ vector of all zeros, and $\mathbf{0}_{n\times p}=\mathbf{0}_n\mathbf{0}_p^T$. The notation $\| \bbX \|_\infty$ and $\| \bbX \|$ stand for the entrywise largest absolute value and the largest singular value of $\bbX$, respectively. For a linear operator $\ccalX$, $\| \ccalX \| := \| \bbX \|$, where $\bbX$ is the matrix representation of $\ccalX$. Otherwise, standard vector and matrix norm notation is used.

\section{Graph Signals and Graph Filters}\label{S:Modeling}

Often, networks have intrinsic value and are themselves the object of study. In 
other occasions, the network defines an underlying notion of proximity, but the object of interest is a signal 
defined over the graph, i.e., data associated with the nodes of the network. This is the matter addressed 
by GSP, where the notions of, e.g., frequency and filtering (reviewed next) are 
extended to signals supported on graphs~\cite{SandryMouraSPMag,EmergingFieldGSP}.
\vspace{1mm}\newline
\noindent\textbf{Graph signals and graph-shift operator.} Let $\mathcal{G}$ denote a directed graph with a set of nodes $\mathcal{N}$ (with cardinality $N$) and a set of links $\mathcal{E}$, if $i$ is connected to $j$ then $(i,j)\in\mathcal{E}$. Since $\ccalG$ is directed, local connectivity is captured by the set $\ccalN_i:=\{j\;|(j,i)\in\mathcal{E}\}$ which stands for the (incoming) neighborhood of $i$. For any given $\mathcal{G}$ we define the adjacency matrix $\bbA\in\mathbb{R}^{N\times N}$ as a sparse matrix with non-zero elements $A_{ji}$ if and only if $(i,j)\in\ccalE$. The value of $A_{ji}$ captures the strength of the connection from $i$ to $j$. 

The focus of the paper is on analyzing and modeling (graph) signals defined on $\mathcal{N}$. These signals can be represented as vectors $\mathbf{x}=[x_1,...,x_N]^T \in  \mathbb{R}^N$, where $x_i$ represents the value of the signal at node $i$. Since the vectorial representation does not account explicitly for the structure of the graph,  $\mathcal{G}$ can be endowed with the so-called \emph{graph-shift operator} $\mathbf{S}$ \cite{SandryMouraSPG_TSP13,SandryMouraSPG_TSP14Freq}. The shift $\mathbf{S}\in\mathbb{R}^{N\times N}$ is a matrix whose entry $S_{ji}$ can be non-zero only if $i=j$ or if $(i,j)\in\mathcal{E}$. The sparsity pattern of the matrix $\bbS$ captures the local structure of $\ccalG$, but we make no specific assumptions on the values of its non-zero entries.
The intuition behind $\mathbf{S}$ is to represent a linear transformation that can be computed locally at the nodes of the graph. More rigorously, if $\mathbf{y}$ is defined as $\mathbf{y}=\mathbf{S}\mathbf{x}$, then node $i$ can compute $y_i$ as linear combination of the signal values $x_j$ at node $i$'s neighbors $j\in \mathcal{N}_i$. For example, one can think of an individual's opinion formation process as one of weighing in the 
views of close friends regarding the subject matter.
Typical choices for $\mathbf{S}$ are the adjacency matrix $\bbA$ \cite{SandryMouraSPG_TSP13,SandryMouraSPG_TSP14Freq}, and the graph Laplacian \cite{EmergingFieldGSP}. We assume henceforth that $\bbS$ is diagonalizable, so that $\bbS=\bbV\bbLambda\bbV^{-1}$ with $\bbLambda\in\mathbb{C}^{N\times N}$ being diagonal. In particular, $\bbS$ is diagonalizable when it is normal, i.e., it satisfies $\bbS\bbS^H=\bbS^H\bbS$. In that case we have that $\bbV$ is unitary, which implies $\bbV^{-1}=\bbV^{H}$, and leads to the decomposition $\bbS=\bbV\bbLambda\bbV^H$.
\vspace{1mm}\newline
\noindent\textbf{Graph filters as models of network diffusion processes.} The shift $\bbS$ can be used to define linear graph-signal \emph{operators} of the form
\begin{eqnarray}\label{E:Filter_input_output_time}
	&\mathbf{H}:=\sum_{l=0}^{L-1}h_l \mathbf{S}^l&
\end{eqnarray}
which are called \emph{graph filters}~\cite{SandryMouraSPG_TSP13}. For a given input $\bbx$, the output of the filter is simply $\bby=\bbH\bbx$. The coefficients of the filter are collected into $\mathbf{h}:=[h_0,\ldots,h_{L-1}]^T$, with $L-1$ denoting the filter degree. Graph filters are of particular interest because they represent linear transformations that can be implemented in a distributed fashion~\cite{EUSIPCO_our_interp_2015,ssamar_distfilters_allerton15}, e.g., with
$L-1$ successive exchanges of information among neighbors. 

Graph filters can be used to model linear diffusion dynamics that depend on the network topology. Formally, the signal at node $i$ during the step $(l+1)$ of a linear diffusion process in $\ccalG$ can be written as
\begin{equation}
\label{E:local_diffusion_singlenode}
x_i^{(l+1)}=\alpha_{ii} x_{i}^{(l)} + \textstyle \sum_{j\in \ccalN_i} \alpha_{ij} x_{j}^{(l)}
\end{equation}
where $\alpha_{ij}$ are the diffusion coefficients. Leveraging the GSP framework, \eqref{E:local_diffusion_singlenode} is equivalent to writing that the graph signal at iteration $l+1$ is the shifted version of the signal at the previous iteration $\bbx^{(l+1)}=\bbS\bbx^{(l)}$, 
where the entries of the shift operator $\bbS$ are  $S_{ij}=\alpha_{ij}$ if either $i=j$ or $(j,i) \in \ccalE$, and $S_{ij}=0$ otherwise.
Notice that if, for example, we set $\bbS=\bbI_N - \beta \bbL$ and say that the signal of interest is $\bby := \bbx^{(\infty)}$, then $\bby$ solves the heat diffusion equation. However, more complex diffusion dynamics, such as $\bby=\Pi_{l=0}^{\infty}(\bbI_N-\beta_l \bbS)\bbx^{(0)}$ and $\bby=\sum_{l=0}^{\infty}\gamma_l \bbx^{(l)}=\sum_{l=0}^{\infty}\gamma_l \bbS^l\bbx^{(0)}$, could also be of interest. 

According to the previous discussion, it is apparent that the steady-state signal $\bby$ generated by a diffusion process can be viewed as the output of a graph filter $\bbH=\sum_{l=0}^{N-1}h_l \bbS^l$ with input $\bbx^{(0)}$. Note also that the Cayley-Hamilton theorem guarantees that the aforementioned
infinite-horizon processes can be equivalently described by a filter of degree $N-1$.
\vspace{1mm}\newline
\noindent\textbf{Frequency domain representation.} Leveraging the spectral decomposition of $\bbS$, graph filters and signals can be represented in the frequency domain.
To be precise, let us use the eigenvectors of $\bbS$ to define the $N\times N$ matrix $\bbU:=\bbV^{-1}$, and the eigenvalues of $\bbS$ to define the $N\times L$ Vandermonde matrix  $\bbPsi$, where $\Psi_{ij}:=(\Lambda_{ii})^{j-1}$.
Using these conventions, the frequency representations of a \emph{signal} $\bbx$ and of a \emph{filter} $\bbh$ are defined as $\widehat{\bbx}:=\bbU\bbx$ and $\widehat{\bbh}:=\bbPsi\bbh$, respectively~\cite{SandryMouraSPG_TSP14Freq}. 
Exploiting such representations, the output $\bby\!=\!\bbH\bbx$ of a graph filter in the frequency domain is given by
\begin{equation}\label{E:Filter_input_output_freq}
	\widehat{\bby}=\diag\big(\bbPsi\bbh\big)\bbU \bbx=\diag\big(\widehat{\bbh}\big)\widehat{\bbx}=
	\widehat{\bbh}\circ\widehat{\bbx}.
\end{equation}
Identity \eqref{E:Filter_input_output_freq} is the counterpart of the celebrated convolution theorem for temporal signals, and follows from $\mathbf{H}=\bbV\big(\sum_{l=0}^{L-1}h_l \boldsymbol{\Lambda}^l\big)\bbU$ [cf. \eqref{E:Filter_input_output_time}] and $\sum_{l=0}^{L-1}h_l \boldsymbol{\Lambda}^l=\diag{(\bbPsi\bbh)}$; see e.g., \cite{SSGMAMAR_camsap15} for a detailed derivation.
To establish further connections with the time domain, let us consider the directed cycle graph whose adjacency matrix $\mathbf{A}_{dc}$ is zero, except for entries $A_{ij}=1$ whenever $i=\mymod(j)+1$, where
$\mymod(x)$ denotes the modulus (remainder) obtained after dividing $x$ by $N$. If $\bbS=\bbA_{dc}$, one can verify that: i) $\bby=\bbH\bbx$ can be found as the circular convolution of $\bbh$ and $\bbx$, and ii) both $\bbU$ and $\boldsymbol{\Psi}$ correspond to the Discrete Fourier Transform (DFT) matrix. Interestingly, while in the time domain $\bbU=\boldsymbol{\Psi}$, this is not true for general graphs.

\section{Blind Identification of Graph Filters}\label{S:BlindIdent}

%
\begin{figure}
\centering
		\includegraphics[width=0.45\textwidth]{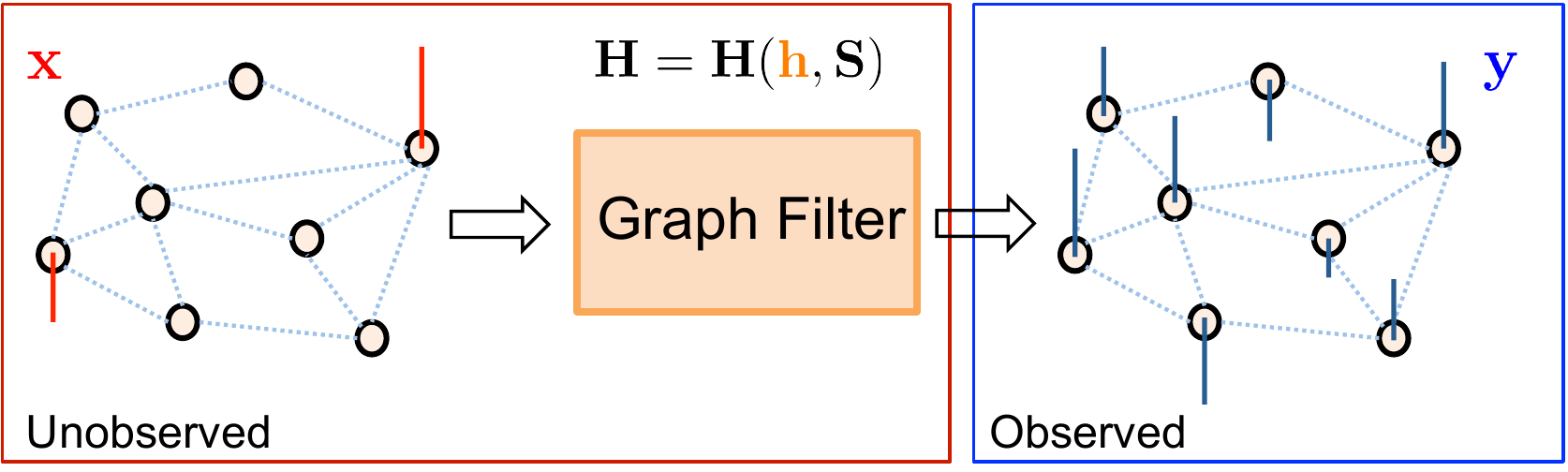} \caption{Setup for the blind graph-filter identification problem.
			Given a graph signal $\bby$ modeled as the output of a graph filter, the goal is to recover the vector of filter coefficients $\bbh$,
			and the input signal $\bbx$ that is assumed to be sparse.}
		\label{F:system_setup}
\end{figure}
The concepts introduced in the previous section can be used to formally state the problem. 
For given shift operator $\bbS$ and filter degree $L-1$, suppose that we observe the output signal $\bby=\bbH\bbx$ [cf. \eqref{E:Filter_input_output_time}], 
where $\bbx$ is \emph{sparse} having at most $S\ll N$ non-zero entries. For future reference introduce the $\ell_0$
(pseudo) norm $\|\bbx\|_0:=|\text{supp}(\bbx)|$, where the support of $\bbx$ is $\text{supp}(\bbx):=\{i \, | \, x_i\neq 0\}$ and hence $\|\bbx\|_0\leq S$.
The present paper deals with \emph{blind identification} of the graph filter (and its input signal),
which amounts to estimating sparse $\bbx$ and the filter coefficients $\bbh$ from the observed output signal $\bby$; see Fig. \ref{F:system_setup}. This problem is a natural extension to graphs of classical blind system identification, or blind deconvolution of signals in the temporal or spatial domains.
\begin{remark}[Sparse input]\normalfont
	Sparsity in $\bbx$ is well-motivated due to its practical 
	relevance and modeling value -- network signals such as $\bby$ are oftentimes the diffused version of few localized sources, hereby indexed by $\text{supp}(\bbx)$. In addition, the non-sparse formulation with $S=N$ is ill-posed, since the number of unknowns $N+L$ in $\{\bbx,\bbh\}$ exceeds the number of observations $N$ in $\bby$. Alternatively, a low-dimensional subspace model for $\bbx$ could be also adopted to effectively reduce the degrees of freedom in the problem~\cite{SSGMAMAR_camsap15}.
\end{remark}
%
Given the observed filtered output $\bby$, one can obtain its frequency-domain representation $\widehat{\bby}=\bbU\bby=\diag\big(\bbPsi\bbh\big)\bbU \bbx$ [cf. \eqref{E:Filter_input_output_freq}] and state
the blind graph-filter identification problem as the following feasibility problem
\begin{align}\label{E:vectorbiconvex_zeronorm_min}
	\text{find}&\:\,\,\{\bbh,\bbx\}\nonumber\\
	\text{s. to }&\:\,\, \widehat\bby= \diag\big(\bbPsi\bbh\big)\bbU \bbx,\:\,\, \|\bbx\|_{0}\leq S.
\end{align}
In other words, the goal is to find the solution to a set of bilinear equations subject to a sparsity
constraint in $\bbx$. 

\subsection{Lifting the Bilinear Constraints}\label{Ss:Lifting}

While very natural, \eqref{E:vectorbiconvex_zeronorm_min} is in fact
a difficult problem due to the non-convex $\ell_0$-norm as well as the bilinear constraints. 
To deal with the latter, it is convenient to rewrite the first constraint in \eqref{E:vectorbiconvex_zeronorm_min} as 
\begin{equation}\label{E:Filter_input_output_time_khatri}
	\widehat\bby=\big(\bbPsi^T\odot\bbU^T\big)^T\text{\normalfont vec}\big(\bbx\bbh^T\big)
\end{equation}
where $\odot$ denotes the Khatri-Rao (i.e., columnwise Kronecker) product, and $\text{\normalfont vec}(\cdot)$
is the matrix vectorization operator. To establish \eqref{E:Filter_input_output_time_khatri}, let $\bbu_i^T$ and $\bbpsi_i^T$ denote the $i$-th rows of $\bbU$ and $\bbPsi$, respectively. It follows from \eqref{E:Filter_input_output_freq} that $\widehat{y}_i=(\bbpsi_i^T\bbh)(\bbu_i^T\bbx)=\big(\bbpsi_i^T\otimes\bbu_i^T\big)\text{vec}\big(\bbx\bbh^T\big)$, where $\otimes$ denotes the Kronecker product. Upon stacking the entries $\widehat{y}_i$ to form $\widehat{\bby}$, the result follows by identifying $\bbpsi_i^T\otimes\bbu_i^T$ with the $i$-th row of $\big(\bbPsi^T\odot\bbU^T\big)^T$.

While \eqref{E:Filter_input_output_time_khatri} confirms that the filtered graph signal $\widehat\bby$ is a \textit{bilinear} function of $\bbx$ and $\bbh$, it also shows that $\widehat\bby$ is a \emph{linear}
combination of the entries of the lifted rank-one, outer-product matrix $\bbZ:=\bbx\bbh^T\in\reals^{N\times L}$. In other words, there exists a linear mapping $\ccalM:\reals^{N\times L}\mapsto\mathbb{C}^N$ such that
$\widehat \bby = \ccalM(\bbZ)$. Note that $\ccalM$ can be expressed
in terms of a matrix multiplication with $\bbM:=\big(\bbPsi^T\odot\bbU^T\big)^T\in \mathbb{C}^{N\times LN}$, since $\widehat\bby = \bbM\text{vec}(\bbZ)$ as per \eqref{E:Filter_input_output_time_khatri}. 
In addition to being of rank one, note that the sparsity in $\bbx$ renders $\bbZ$ row-wise sparse, i.e., rows $\bbz_i^T$ indexed by $\{1,\ldots,N\}\setminus\text{supp}(\bbx)$ are identically zero.
Building on the ideas in~\cite{ahmed2014BlindDeconvConvex,LingBiConvexCS},
one can thus pose the blind graph-filter identification problem as a \emph{linear} inverse problem, where the goal is to recover a row-sparse, rank-one
$N\times L$ matrix $\bbZ$ from observations $\widehat\bby=\ccalM(\bbZ)$.  To this end, a natural formulation to tackle such inverse problem is
\begin{align}\label{E:Rank_min}
	\min_{\bbZ}&\:\,\,\mathrm{rank}(\bbZ)\nonumber\\
	\text{s. to }&\:\,\, \widehat\bby\!=\!\big(\bbPsi^T\!\odot\bbU^T\big)^T\text{vec}\big(\bbZ\big),\:\,\,\|\bbZ\|_{2,0}\leq S
\end{align}
where $\|\bbZ\|_{2,0}$ is equal to the number of non-zero rows of $\bbZ$.

A basic question is whether \eqref{E:Rank_min} is equivalent to the original blind identification problem. 
To give a rigorous answer, some definitions are introduced next. For a given matrix $\bbU$, $\mathrm{spark}(\bbU)$ is the smallest number $n$ such that there exists a subgroup of $n$ columns from $\bbU$ that
are linearly dependent~\cite{donohospark}. Given a set of row indices $\ccalI$, define the complement set of indices $\ccalI^c := \{1, \ldots, N \} \backslash \ccalI$ and the matrix $\bbU_\ccalI$ formed by the rows of $\bbU$ indexed by $\ccalI$. Moreover, for a given graph-shift operator $\bbS$ -- fixed $\bbV$, $\boldsymbol{\Psi}$, and $\bbU$ -- define the set $\ccalO_{\widehat\bby}$ of matrix minimizers of \eqref{E:Rank_min} as a function of $\widehat\bby$. Then, the following result on the validity of the matrix problem formulation in \eqref{E:Rank_min} holds.
\begin{myproposition}\label{P:validity_problem_formulation}
	Let $\ccalI_{S}$ be a set of row indices such that $\mathrm{spark}(\bbU_{\ccalI_{S}}) \leq S$. Then, the set of minimizers of \eqref{E:Rank_min}, satisfies 
	\vspace{-0.1in}
	\begin{eqnarray}\label{E:equality_of_solution_sets}
		&\ccalO_{\widehat\bby} = \big\{\bbx \bbh^T \, \big| \,\widehat\bby = \bbU\sum_{l=0}^{L-1}h_l \mathbf{S}^l \bbx,\:\:\|\bbx\|_0\leq S \big\}&
	\end{eqnarray}
	for any $\widehat\bby$ if and only if
	\begin{equation}\label{E:condition_cardinality_lambdas}
		\min_{\ccalI_{S}} \big| \{ \lambda_i \}_{i \in \ccalI^c_{S}} \big| > L-1.
	\end{equation}
\end{myproposition}
\begin{myproof}
	If we show that \eqref{E:condition_cardinality_lambdas} is violated if and only if there exists a rank-one matrix $\bbZ = \bbx \bbh^T$ such that $\big(\bbPsi^T\odot\bbU^T\big)^T\text{vec}\big(\bbZ\big) = \mathbf{0}_N$ and $\|\bbZ\|_{2,0}\leq S$, then Corollary 1 in \cite{choudhary2014identifiability} completes the proof. The above system of homogeneous equations can be written as $(\bbpsi_i^T\bbh) ( \bbu_i^T \bbx) = 0$ for $i = 1, \ldots, N$, where $\bbpsi_i^T$ denotes the $i$-th row of $\bbPsi$ and similarly for $\bbU$. Since $\mathrm{spark}(\bbU_{\ccalI_{S}}) \leq S$, there exists $\bbx \neq \mathbf{0}_N$ with $\|\bbx\|_0\leq S$ such that $(\bbpsi_i^T\bbh) ( \bbu_i^T \bbx) = 0$ holds for $i \in \ccalI_{S}$. Exploiting the Vandermonde structure of $\bbPsi$, it follows that $\bbh \neq \mathbf{0}_L$ satisfying the equality for $i \in \ccalI^c_{S}$ can be found if and only if \eqref{E:condition_cardinality_lambdas} is violated.
\end{myproof}

Ideally, when solving \eqref{E:Rank_min} for some output $\widehat\bby$ one should recover the set of outer products of all possible combinations of sparse inputs $\bbx$ and filter coefficients $\bbh$ that can give rise to such output [cf. \eqref{E:equality_of_solution_sets}]. This is not true in general \cite[Theorem 1]{choudhary2014identifiability}, however, Proposition \ref{P:validity_problem_formulation} states conditions on the graph-shift operator [cf. \eqref{E:condition_cardinality_lambdas}] for the desired equivalence to hold. For the particular case of the directed cycle graph, we may select the support of $\bbx$ so that every choice of $S$ rows of $\bbU$ forms a full-rank matrix. Consequently, the cardinality in \eqref{E:condition_cardinality_lambdas} is equal to $N-S+1$ entailing the following corollary.

\begin{mycorollary}\label{C:directed_cycle}
	If $\bbS = \bbA_{dc}$ and the support of $\bbx$ consists of either $S$ adjacent or $S$ equally spaced nodes, then \eqref{E:equality_of_solution_sets} holds if and only if $N > L + S - 2$.
\end{mycorollary}

Notice that condition \eqref{E:condition_cardinality_lambdas} does not guarantee that the solution of \eqref{E:Rank_min} is unique, but rather that the outer product of the desired sparse signal and filter coefficients is contained in $\ccalO_{\widehat\bby}$. For instance, uniqueness necessarily requires $\text{rank}(\bbPsi)=L$, otherwise there is no hope to recover the actual $\bbh$ from observations $\widehat\bby$ [cf. \eqref{E:vectorbiconvex_zeronorm_min}]. 

\subsection{Algorithmic Approach via Convex Relaxation}\label{Ss:Algorithm}

Albeit natural, problem \eqref{E:Rank_min} is challenging since both the rank and the $\ell_0$-norm are in general NP-hard to optimize; see e.g.,~\cite{recht2010RankMinimization}. Over the last decade or so, convex relaxation approaches to tackle rank and/or sparsity minimization problems have enjoyed remarkable success, since they oftentimes entail no loss in optimality. The nuclear norm $\|\bbZ\|_*=\sum_k\sigma_k(\bbZ)$, where $\sigma_k(\bbZ)$ denotes the $k$-th singular value of $\bbZ$, is typically adopted as a convex surrogate to
$\mathrm{rank}(\bbZ)$~\cite{fazel2001RankMinimization,recht2010RankMinimization}. Likewise, the $\ell_{2,1}$ mixed norm  $\|\bbZ\|_{2,1}:=\sum_{i=1}^N\|\bbz_i^T\|_2$ is the closest convex approximation of $\|\bbZ\|_{2,0}$~\cite{tropp06tit}. With $\tau$ denoting a tuning parameter to control the rank versus row-sparsity tradeoff, a convex heuristic is to solve
\begin{align}\label{E:Nuclear_l21_norms_min}
	\min_{\bbZ}&\:\,\,\|\bbZ\|_*+\tau\|\bbZ\|_{2,1}\nonumber\\
	\text{s. to } &\:\,\, \widehat \bby=\big(\bbPsi^T\odot\bbU^T\big)^T\text{vec}\big(\bbZ\big)
\end{align}
hoping that the optimal solution is of rank one and has $S$ non-zero rows, so that we can recover $\bbx$ and $\bbh$ up to scaling. 

Recovery of simultaneously low-rank and row-sparse matrices from noisy compressive measurements
was also considered in~\cite{LowRankSparseHyperspectralICASSP12} for hyperspectral image reconstruction.
Recent theoretical results on recovery of simultaneously structured matrix models suggest
that minimizing only $\|\bbZ\|_{1}$  could as well suffice~\cite{oymak2015SimultStruct}; see also~\cite{LingBiConvexCS}, the discussion at the end of this section and the performance guarantees in Section \ref{S:Recovery}. Being convex, \eqref{E:Nuclear_l21_norms_min} is computationally appealing, in fact off-the-shelf interior point solvers are available.
Customized scalable algorithms for large-scale graphs can be developed to minimize the composite, non-differentiable cost in \eqref{E:Nuclear_l21_norms_min}. For instance
the solver implemented to run the numerical tests in Section \ref{S:Simulations} leverages the
alternating-direction method of multipliers (ADMM)~\cite{BoydADMM}; see also~\cite{LowRankSparseHyperspectralICASSP12} for a related proximal-splitting algorithm. \vspace{1mm}\newline
\noindent\textbf{Refinement via iteratively-reweighted optimization.}
Instead of substituting $\|\bbZ\|_{2,0}$ in \eqref{E:Rank_min}
by its closest convex approximation, namely
$\|\bbZ\|_{2,1}$, letting the surrogate function to be
non-convex can yield tighter approximations, and potentially improve the statistical 
properties of the estimator. 
In the context of sparse signal recovery for instance, the
$\ell_0$ norm of a vector was surrogated in~\cite{candes_l0_surrogate} by the logarithm
of the geometric mean of its elements.

Building on this last idea, 
consider replacing $\|\bbZ\|_{2,1}$ in \eqref{E:Nuclear_l21_norms_min}  with $\sum_{i=1}^N\log(\|\bbz_i^T\|_2+\delta)$, where $\delta$ is a small positive constant. Since the new surrogate term is concave, the overall 
minimization problem is non-convex and admittedly more complex to solve
than \eqref{E:Nuclear_l21_norms_min}. With $k$ denoting iterations, 
local methods based on iterative linearization of
$\log(\|\bbz_i^T\|_2+\delta)$ around the current iterate
$\bbz_i^T(k)$, can be adopted to minimize the resulting non-convex cost. Skipping details that
can be found in~\cite{candes_l0_surrogate}, application of the majorization-minimization 
technique leads to an \emph{iteratively-reweighted} version of \eqref{E:Nuclear_l21_norms_min}, namely
solve for $k=0,1,\dots$
\begin{align}\label{E:reweighted_problem}
	\min_{\bbZ}&\:\,\,\|\bbZ\|_*+ \sum_{i=1}^N w_i(k)\|\bbz_i^T\|_2\nonumber\\
	\text{s. to }&\:\,\,  \widehat\bby=\big(\bbPsi^T\odot\bbU^T\big)^T\text{vec}\big(\bbZ\big)
\end{align}
with weights $w_i(k):=\tau/\left(\|\bbz_i^T(k-1)\|_2+\delta\right)$. 
If the value of $\|\bbz_i^T(k \! -\!1)\|_2$ is small, then in the next iteration the regularization
term $w_i(k)\|\bbz_i^T\|_2$ has a large weight,
promoting shrinkage of that entire row vector to zero. Numerical tests in Section \ref{S:Simulations} suggest that few iterations of the iteratively-reweighted procedure suffice to yield improved recovery of $\{\bbx,\bbh\}$, when compared to \eqref{E:Nuclear_l21_norms_min}.\vspace{1mm}\newline
\noindent\textbf{Blind identification via linear programming.} Since schemes aimed at finding sparse matrices can also lead to low-rank solutions \cite{LingBiConvexCS,oymak2015SimultStruct}, the last proposed relaxation adopts a single-structure enforcing criterion under which, as in SparseLift~\cite{LingBiConvexCS}, only the $\ell_1$-norm of $\bbZ$ is minimized 
\begin{align}\label{E:l1_norms_min}
\min_{\bbZ}&\:\,\,\|\bbZ\|_{1}\nonumber\\
\text{s. to } &\:\,\,  \widehat\bby=\big(\bbPsi^T\odot\bbU^T\big)^T\text{vec}\big(\bbZ\big).
\end{align}
The above optimization is a linear program that, if needed, can be modified to accommodate an iteratively-reweighted counterpart. While we know $\bbZ$ is low-rank and row-wise sparse, we admittedly relaxed the structural constraints
and only encouraged $\bbZ$ to be entry-wise sparse, with no specific pattern preferred a fortiori. The reason for considering the simpler formulation in \eqref{E:l1_norms_min} is threefold. First, the absence of $\|\bbZ\|_*$ circumvents the need to perform a singular value decomposition (SVD) per iteration, as is customary with nuclear-norm
minimization. Second,  \eqref{E:l1_norms_min}  eliminates the burden of selecting an adequate tuning parameter $\tau$ in 
\eqref{E:Nuclear_l21_norms_min}. Third, as
we establish in Section \ref{S:Recovery}, under some conditions the simplification in \eqref{E:l1_norms_min} is enough to uniquely recover $\bbZ=\bbx\bbh^T$ with high probability.

\begin{remark}[Noisy and partial observations]
\normalfont The proposed relaxations can be easily modified to account for noisy or partial observations of the graph signal $\bby$. Following the standard approach for sparse recovery problems, when the observations $\bby$ are noisy, it suffices to rewrite the filter output constraint 
as
$\|\bbV\widehat{\bby}-\bbV(\bbPsi^T\odot\bbU^T)^T\text{vec}(\bbZ)\|_2^2\leq \epsilon^2$
, where the specific norm and value of $\epsilon^2$ will depend on the observation noise model. Moreover, it is not uncommon to encounter graph-based settings where one measures $\bby$ in a subset of nodes only. This could happen
because it is impossible to access parts of the network, or due to intentional sampling with the goal of reducing overall processing complexity. Accordingly, suppose that $C\leq N$ and define the partially observed signal $\bby_c\in \reals^{C}$ as 
$\bby_c:=\bbC\bby=\bbC\bbV\widehat{\bby}$,
with $\bbC$ being a sampling matrix formed by a subset of rows of the $N\times N$ identity matrix. The graph filter output constraint should be now written as $ \bby_c=\bbC\bbV(\bbPsi^T\odot\bbU^T)^T\text{vec}(\bbZ)$, so that 
the matrix $\bbC\bbV$ 
is incorporated into the linear mapping $\ccalM(\bbZ)$. Because the input signal $\bbx$ is assumed to be sparse, it may still be feasible to recover $\{\bbh,\bbx\}$
from partial observations $\bby_c$; see also the numerical tests in Section \ref{S:Simulations}.
\end{remark}

\subsection{Multiple Output Signals}\label{S:MultipleOutputs}

Jointly processing multiple output signals (when available)
can aid the blind identification task, and this is the subject of the present section.
Suppose now that we have access to a collection of $P$ (possibly time-indexed) output signals $\{\bby_p\}_{p=1}^P$, each one corresponding to a different sparse input $\bbx_p$ fed to the \emph{common} graph filter $\bbH$ we wish to identify. Although each of the $P$ identification problems could be solved separately (and naively) as per Section \ref{S:BlindIdent}, the recovery performance can be improved by tackling them jointly.

While extending the feasibility problem in \eqref{E:vectorbiconvex_zeronorm_min} to this new setup is straightforward [each output gives rise to a couple constraints as in \eqref{E:vectorbiconvex_zeronorm_min}], generalizing the formulation in \eqref{E:Nuclear_l21_norms_min} requires more work. To this end, consider the $NP\times 1$ supervector of stacked output signals $\tilde{\bby}:=[\widehat\bby_1^T,...,\widehat\bby_P^T]^T$, and likewise for the unobserved inputs $\tilde{\bbx}:=[\bbx_1^T,...,\bbx_P^T]^T$. Next, introduce the unknown rank-one matrices $\bbZ_p:=\bbx_p\bbh^T$, $p=1,...,P$, and stack them: (i) vertically in $\tilde{\bbZ}_{v}:=[\bbZ_1^T,...,\bbZ_P^T]^T=\tilde{\bbx}\bbh^T\in\reals^{NP\times L}$; and (ii) horizontally in $\tilde{\bbZ}_h:=[\bbZ_1,...,\bbZ_P]\in\reals^{N\times PL}$. Note that $\tilde{\bbZ}_{v}$ is a rank-one matrix. Further, when all 
$\bbx_p$ share a common support, then so will all the row-sparse matrices $\bbZ_p$ (and hence $\tilde{\bbZ}_h$). These observations motivate the following convex formulation [cf. \eqref{E:Nuclear_l21_norms_min}]
\begin{align}
	\label{E:Nuclear_l21_multiple_outputs} \min_{ \{\bbZ_p\}_{p=1}^P } \;\;&\| \tilde{\bbZ}_{v}\|_*+\tau\|\tilde{\bbZ}_{h}\|_{2,1}\\
	\nonumber \text{s. to } \;\;&\tilde{\bby}=\left( \bbI_P \otimes \Big( \big(\bbPsi^T\odot\bbU^T\big)^T\Big)\right) \text{vec}\big(\tilde{\bbZ}_{h}\big)
\end{align}
where all $P$ lifted bilinear constraints have been compactly expressed in terms of $\tilde{\bby}$ and $\text{vec}\big(\tilde{\bbZ}_{h}\big)$ using a Kronecker product. 

When the sparse support is not the same for all $\bbx_p$, matrix $\tilde{\bbZ}_{h}$ is not row-sparse. In that case, $\|\tilde{\bbZ}_{h}\|_{2,1}$ in \eqref{E:Nuclear_l21_multiple_outputs} must be replaced with $\sum_{p=1}^P \|\bbZ_p\|_{2,1}$, possibly adjusting individual tuning parameters $\tau_p$ per signal. Either way, an efficient ADMM solver can be implemented for the
multiple signal setting as well, and extensive numerical tests indicated that
iteratively-reweighing as in Section \ref{Ss:Algorithm} can yield markedly improved 
recovery performance (cf. Section \ref{S:Simulations}).


\section{Exact Recovery via Convex Optimization}\label{S:Recovery}

Here we show that under some technical conditions, both the filter coefficients
$\bbh$ as well as the sparse input signal $\bbx$ in \eqref{E:Filter_input_output_time_khatri} can be exactly recovered 
by solving the convex problem \eqref{E:l1_norms_min}. 
Assumptions delineating the analysis' scope are first outlined in Section \ref{Ss:Scope}, after which the main
result is formally stated in Section \ref{Ss:Results} followed by a discussion of the recovery conditions and their dependence
on the graph. The proof follows closely the ideas in~\cite{LingBiConvexCS} and the key steps are given in Appendix \ref{App:proof_the_noise_free}, with an emphasis on the novel aspects introduced by the GSP context dealt with here. 

\subsection{Assumptions and Scope of the Analysis}\label{Ss:Scope}

Two main assumptions are made to facilitate the analysis.\vspace{1.5mm}\\
%
\noindent\textbf{(as1)} The graph-shift operator $\bbS$ is normal, i.e., it satisfies $\bbS\bbS^H=\bbS^H\bbS$, and its eigenvalues are all distinct.
\vspace{1.5mm}
\newline
\noindent\textbf{(as2)} The frequency representation of the observed graph~signal $\bby$ adheres to the model
$\widehat{\bby}=\diag(\bbPsi\bbh) \tilde{\bbU} \bbx$, where $\tilde{\bbU}$ is a random $N \! \times \! N$ matrix obtained by concatenating $N$~rows sampled independently and uniformly with replacement from~$\bbU$.
\vspace{1.5mm}
\newline
Under (as1) $\bbV$ is unitary, which implies $\bbU:=\bbV^{-1}=\bbV^{H}$ and leads to the decomposition $\bbS=\bbV\bbLambda\bbV^H$. Normality is for instance satisfied when $\ccalG$ is undirected and the graph-shift operator is chosen to be the adjacency matrix or the graph Laplacian. Furthermore, all the eigenvalues of $\bbS$ being distinct ensures that matrix $\bbPsi$ is full rank independently of $L$, which is required for uniqueness as discussed in the end of Section \ref{Ss:Lifting}. Under (as1), we can assume that $\bbPsi^H \bbPsi = \bbI_L$ without loss of generality. To see this, consider, e.g., the SVD of $\bbPsi =  \bbP \bbSigma \bbR^H$ and rewrite the frequency response of the filter as $\bbPsi \bbh = \bbP \bbSigma \bbR^H \bbh : = \bbP \bbh'$, where $\bbP$ satisfies $\bbP^H \bbP = \bbI_L$ by definition and $\bbh$ can be recovered from $\bbh'$ due to the full-rank condition of $\bbPsi$. Consequently, in the statement of Theorem~\ref{T:noise_free} and its proof we assume that $\bbPsi^H \bbPsi = \bbI_L$.

Regarding the probabilistic model for the observations in (as2), this type of models are customary towards establishing recovery guarantees in the context of, e.g., 
compressed sensing and low-rank matrix completion~\cite{math_cs2013book}, or even blind deconvolution of temporal signals~\cite{ahmed2014BlindDeconvConvex,LingBiConvexCS}. 
For instance, instrumental to the proof arguments in~\cite{LingBiConvexCS},  is that rows of the matrix representation of operator $\ccalM$ are independent. This way, one can bring to bear matrix Bernstein inequalities to bound the norm of relevant operators constructed from sums of these rows~\cite{TroppMatrixConcentration}.  A direct consequence
is that the recovery results obtained here are probabilistic in nature, namely Theorem \ref{T:noise_free} asserts that $\{\bbx,\bbh\}$ can be recovered with high probability over the measure induced by the aforementioned matrix
randomization procedure, which can also be interpreted as inducing a particular ensemble of random graphs.  All in all, tractability is the main reason behind (as2), which resembles the random Fourier model in~\cite{LingBiConvexCS} but is more general since rows are sampled from a unitary matrix $\bbU$ -- not necessarily the DFT matrix.

In any case, we would like to stress that the focus here is only on establishing that a convex relaxation
can succeed for blind identification of graph filters, and that the graph structure plays a key role on the recovery performance. 
We are not after the tightest guarantees, and the success probability bounds obtained are admittedly loose. In fact, Theorem \ref{T:noise_free}
deals with recovery of $\{\bbx,\bbh\}$ using the simplified convex formulation \eqref{E:l1_norms_min}, despite of the fact that \eqref{E:Nuclear_l21_norms_min} exhibits slightly better performance than \eqref{E:l1_norms_min} in practice; see also the numerical tests in Section \ref{S:Simulations}. Nevertheless, in theory both schemes are equivalent (at least order-wise)~\cite{oymak2015SimultStruct}, while the optimality conditions and corresponding construction of dual certificates for \eqref{E:l1_norms_min} are markedly simpler. 

\subsection{Main Result}\label{Ss:Results}

Given an arbitrary matrix $\bbA \in \mathbb{C}^{M \times N}$ and a positive integer $k \in \mathbb{N}$, we define the function $\rho_\bbA(k)$ as
\begin{equation}\label{E:def_rho}
	\rho_\bbA(k) := \max_{l \in \{1, \ldots, M\}} \,\,\, \max_{\Omega \in \Omega^N_k}   \| \bba_{l, \, \Omega} \|_2^2
\end{equation}
where $\Omega^N_k$ represents the set of all $k$-subsets of $\{1, \ldots, N\}$, and $\bba_{l, \Omega}$ is the orthogonal projection of the $l$-th row of $\bbA$ onto the index set $\Omega$. In words, $\rho_\bbA(k)$ is  the largest squared-norm of any vector formed by selecting $k$ elements from a row of $\bbA$. This extends the concept of mutual coherence between the basis of Kronecker deltas on the graph and $\bbA$ \cite{Shuman2016vertex_freq}. The above definition allows us to formalize the following main result.

\begin{mytheorem}\label{T:noise_free}
	For a given graph-shift operator $\bbS$, assume that an $S$-sparse graph signal $\bbx_0 \in\reals^N $ when passed through a filter with coefficients $\bbh_0 \in\reals^L$ results in a signal with frequency representation $\widehat{\bby} \in \mathbb{C}^N$ adhering to the model in (as2). Also, denote by $\bbU\in\mathbb{C}^{N\times N}$ and $\bbPsi\in\mathbb{C}^{N\times L}$ the GFT for signals and filters associated with $\bbS$, respectively, where $\bbU$ is normalized such that $\bbU^H \bbU = N \bbI_N$. Define
	\begin{align}\label{E:main_result_recoverability_alpha_1} 
	\alpha := \frac{3 \log(2) \, \left( 120 \frac{\rho_{\bbU}(1)\rho_{\bbPsi}(1) L S}{\rho_{\bbU}(S)\rho_{\bbPsi}(L)} + 8 \sqrt{\frac{\rho_{\bbU}(1)\rho_{\bbPsi}(1) L S}{\rho_{\bbU}(S)\rho_{\bbPsi}(L)}}\right)^{-1}}{{\rho_{\bbU}(S)\rho_{\bbPsi}(L) } \log(4 \gamma \sqrt{2LS}) \! \log(2SN^2)}
	\end{align}
where $\gamma :=\sqrt{2N(\log(2LN)+1)+1}$. Under (as1)-(as2), if $\alpha \geq 1$ then the unique solution to \eqref{E:l1_norms_min} is the rank-one matrix $\bbZ_0 := \bbx_0\bbh_0^T$, with probability at least 
	\begin{equation}\label{E:main_result_recoverability_prob}
		P_{\text{rec}}\geq 1 - N^{-\alpha + 1}.
	\end{equation}
\end{mytheorem}
\begin{myproof}
See Appendix~\ref{App:proof_the_noise_free}.
\end{myproof}

We want to emphasize three differences between the above theorem and \cite[Th. 3.1]{LingBiConvexCS}. First and foremost, Theorem~\ref{T:noise_free} provides probabilistic guarantees of recovery for blind identification in {\textit{arbitrary}} graphs [cf. (as1)], whereas the results in \cite{LingBiConvexCS} only apply for the cases where $\tilde{\bbU}$ is random Fourier or Gaussian distributed. Our generalization is reflected through the function {$\rho_\bbA$}, which as detailed after Lemma~\ref{L:behavior_rho} also provides intuition about which graph topologies favor blind recovery. Secondly, we provide exact expressions for the constants throughout the proof -- some of them embedded in expression \eqref{E:main_result_recoverability_alpha_1}. Accordingly, bounds for the probability of recovery are derived for finite values of $L$, $S$, and $N$, instead of order-wise asymptotic results. Lastly, by using $\rho_{\bbPsi}$ to describe properties of $\bbPsi$, the bounds obtained in Theorem~\ref{T:noise_free} are tighter than those in \cite[Th. 3.1]{LingBiConvexCS}, even for the case where $\tilde{\bbU}$ is random Fourier. 

Leveraging the facts that both $\rho_{\bbU}$ and $\rho_{\bbPsi}$ are non-decreasing functions by definition [cf. \eqref{E:def_rho}], we can lower bound $\alpha$ in \eqref{E:main_result_recoverability_alpha_1} to obtain an alternative expression which is more restrictive but simpler to understand, namely
\begin{equation}\label{E:alternative_alpha}
\!	\alpha_1 := \frac{3 \log(2) / 128}{L S \rho_{\bbU}(S) \rho_{\bbPsi}(L)  \log(4 \gamma \sqrt{2LS}) \! \log(2SN^2)}\leq \alpha.\!
\end{equation}
%
Inspection of \eqref{E:alternative_alpha} clearly shows that the recovery performance depends on $S$ and $L$ through the functions $\rho_{\bbU}$ and $\rho_{\bbPsi}$, respectively. The following lemma characterizes the behavior of these functions.

\begin{mylemma}\label{L:behavior_rho}
The functions $\rho_{\bbU}$ and $\rho_{\bbPsi}$ as defined in \eqref{E:def_rho} satisfy
\begin{align}
S \leq \rho_{\bbU}(S) \leq S \rho_{\bbU}(1), \label{E:behavior_rho_U} \\
L/N \leq \rho_{\bbPsi}(L) \leq L \rho_{\bbPsi}(1) \label{E:behavior_rho_Psi}
\end{align}
for all graph shift $\bbS$. Moreover, \eqref{E:behavior_rho_U} and \eqref{E:behavior_rho_Psi} are satisfied with equalities when $\bbS = \bbA_{dc}$.
\end{mylemma}
\begin{myproof}
{Since \eqref{E:behavior_rho_U} and \eqref{E:behavior_rho_Psi} can be shown using similar arguments, we focus only on proving \eqref{E:behavior_rho_U}.} To show the rightmost inequality in \eqref{E:behavior_rho_U}, we leverage the definition of {$\rho_\bbA$} in \eqref{E:def_rho} to write
\begin{align}
\rho_{\bbU}(S) & \leq \max_{l \in \{1, \ldots, N\}}  \,\, \max_{\Omega \in \Omega^N_S} \,\, S \,\, \max_{i \in \Omega} \,\, | u_{li} |^2 \nonumber \\
& = S  \max_{l \in \{1, \ldots, N\}}  \,\, \max_{i \in \{1, \ldots, N\}} \,\, |u_{li} |^2 = S \rho_{\bbU}(1)
\end{align}
where the first equality follows from the fact that maximizing over all $S$-subsets first and then maximizing over a particular entry $i \in \Omega$ is equivalent to an initial maximization over $i \in \{1, \ldots, N\}$. To show the leftmost inequality in \eqref{E:behavior_rho_U}, we again rely on \eqref{E:def_rho} to write
\begin{align}
\frac{N}{S} \rho_{\bbU}(S) &= \max_{l \in \{1, \ldots, N\}}  \,\, \frac{N}{S} \max_{\Omega \in \Omega^N_S} \,\, \| \bbu_{l, \, \Omega} \|_2^2 \nonumber \\
&\geq \max_{l \in \{1, \ldots, N\}} \| \bbu_l \|_2^2 = N
\end{align}
where the last equality follows from the fact that $\bbU^H \bbU = N \bbI_N$. Finally, whenever $\bbS = \bbA_{dc}$ notice that $\bbU$ is a DFT matrix with unit-magnitude elements and $\bbPsi$ consists of columns from a normalized DFT matrix (with elements of magnitude $1/\sqrt{N}$), thus equalities in \eqref{E:behavior_rho_U} and \eqref{E:behavior_rho_Psi} follow.
\end{myproof}

\begin{figure}
	\centering
\includegraphics[height=0.26\textwidth, width=0.39\textwidth]{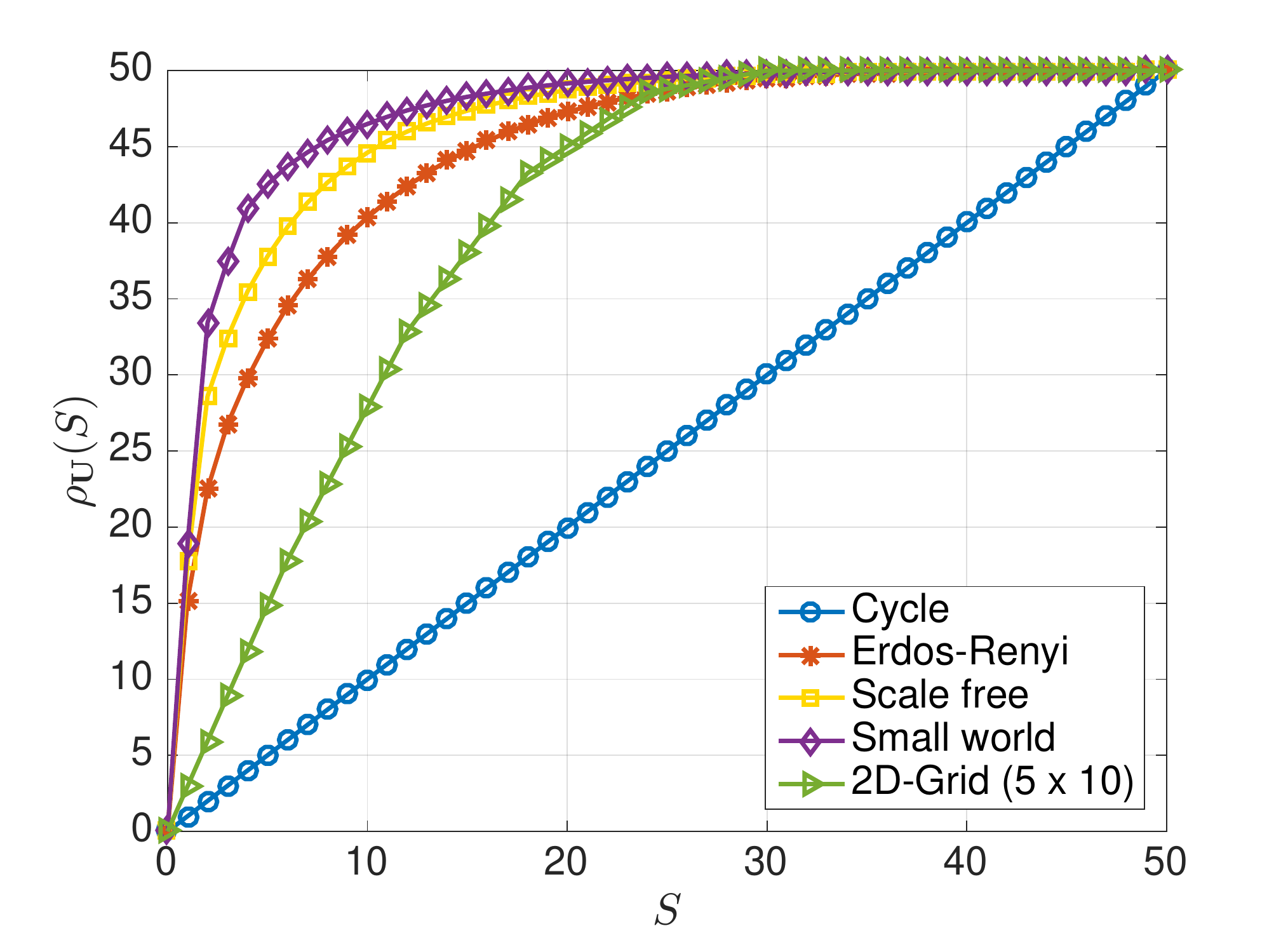}

\vspace{-0.1in}
\caption{\small $\rho_{\bbU}(S)$ for five types of graphs with $N = 50$ nodes. The graph parameters were chosen so that the expected number of edges {is the same for the} three random topologies considered. The values of $\rho_{\bbU}$ reported are the average among 100 realizations. We can see $\rho_{\bbU}$ achieving its lower bound for the directed cycle (cf. Lemma~\ref{L:behavior_rho}).}
\vspace{-0.15in}
\label{F:rho}
\end{figure}

In order to increase the probability of recovery, it is desirable to obtain large values of $\alpha$ [cf.~\eqref{E:main_result_recoverability_alpha_1} and \eqref{E:main_result_recoverability_prob}]. Consequently, functions $\rho_{\bbU}$ and $\rho_{\bbPsi}$ indicate how the recovery performance decreases with increasing $S$ -- number of non-zero entries of the input -- and $L$ -- number of filter coefficients. In particular, the closer $\rho_{\bbU}$ and $\rho_{\bbPsi}$ are to their lower bounds in \eqref{E:behavior_rho_U} and \eqref{E:behavior_rho_Psi}{, the better -- more specifically, the slower the recovery performance deteriorates with increasing $S$ and $L$.} With reference to the theoretical bound in \eqref{E:main_result_recoverability_prob}, Lemma~\ref{L:behavior_rho} implies that blind identification in time (associated with $\bbS = \bbA_{dc}$) corresponds to the most favorable setting for blind identification of graph filters. The behavior of $\rho_{\bbU}$ for different graphs is depicted in Fig.~\ref{F:rho}. 
{Regarding the behavior of $\rho_{\bbPsi}$, since $\bbPsi$ depends not only on the graph but also on the normalization procedure chosen to achieve $\bbPsi^H \bbPsi = \bbI_L$ [cf. discussion after (as1)], the interpretation of $\rho_{\bbPsi}$ and its dependence on the graph structure is more involved.}
Hence, in the ensuing section we limit our numerical analysis to the effect of $\rho_{\bbU}$ on the recovery performance.

\section{Numerical Results}\label{S:Simulations}

\begin{figure*}[t]
	\centering
	\input{figures/num_experiments.tex}	

	\caption{(a)-(d) Rate of recovery of $\bbx_0$ and $\bbh_0$ as a function of $S$ (sparsity in $\bbx_0$) and $L$ (filter length) in 50-node Erd\H{o}s-R\'{e}nyi graphs (top) and a brain graph (bottom) for different recovery algorithms: (a) $\ell_1$ minimization [cf.~\eqref{E:l1_norms_min}]; (b) $\ell_{2,1}$ plus nuclear norm minimization [cf.~\eqref{E:Nuclear_l21_norms_min}]; (c) reweighted $\ell_{2,1}$ plus nuclear norm [cf.~\eqref{E:reweighted_problem}]; and (d) reweighted $\ell_{2,1}$ plus nuclear norm with $P=5$ output observations [cf.~\eqref{E:Nuclear_l21_multiple_outputs}]. (e) Recovery errors for several methods as a function of the number of observations in the brain network for $L=S=3$. Dashed lines represent recovery with noisy observations.}

	\label{F:num_exp}
\end{figure*}

Four types of experiments are conducted to illustrate the performance of our blind graph-filter identification approach. First, we evaluate the effectiveness of the different relaxations proposed in Section \ref{S:BlindIdent} in random and real-world graphs. Second, we compare the performance of our method with alternative approaches when solving a blind identification problem in a brain graph. Third, we assess the sensitivity of recovery with respect to the graph-dependent parameters identified in Section \ref{S:Recovery}. Lastly, we illustrate how our method can be used to identify the sources of contagion in an epidemic model.

In the aforementioned experiments, we solve blind graph-filter identification problems for different graphs $\ccalG$ while varying the parameters $L$, $S$, $P$, and $N$. The obtained signal and filter-coefficient
estimates will be denoted by $\{\tilde{\bbx},\tilde{\bbh}\}$.
In all cases we define the graph-shift operator as the adjacency matrix of $\ccalG$, $\bbS = \bbA$. The ``true'' vectors $\bbx_0$ and $\bbh_0$ are drawn from standard multivariate
Gaussian distributions and are normalized to unit norm. Given $\bbx_0$ and $\bbh_0$, synthetic observations $\bby$ are generated with frequency components given by \eqref{E:Filter_input_output_time_khatri}. The root-mean-square error $\text{RMSE}:=\|\tilde{\bbx}\tilde{\bbh}^T-
\bbx_0\bbh_0^T\|_F$ is adopted as figure of merit to assess recovery performance.

\vspace{0.1cm}
\noindent\textbf{Recovery performance.}
Defining successful recovery when the RMSE is smaller than 0.01, we empirically estimate the successful recovery rate for Erd\H{o}s-R\'{e}nyi graphs ($N=50$, $p=0.1$) \cite{bollobas1998random} as a function of $L$ and $S$ by averaging the success counts over 20 realizations for each parameter combination; see Fig.~\ref{F:num_exp}(a)-(d) (top). We assess the recovery performance for different convex relaxations of increasing effectiveness: (a) $\ell_1$ minimization [cf.~\eqref{E:l1_norms_min}]; (b) $\ell_{2,1}$ plus nuclear norm [cf.~\eqref{E:Nuclear_l21_norms_min}]; (c) reweighted $\ell_{2,1}$ plus nuclear norm [cf.~\eqref{E:reweighted_problem}]; and (d) reweighted $\ell_{2,1}$ plus nuclear norm with $P=5$ output observations [cf.~\eqref{E:Nuclear_l21_multiple_outputs}].
As expected, the difficulty of the problem increases when either $L$ or $S$ increase, depicted in the figures by the darker area around the bottom-right corners. Moreover, when going from one figure to the next, the growing white regions portray the benefits of leveraging additional signal structure in the algorithms. In particular, when moving from (a) to (b), we observe the benefit of incorporating the row-sparse and low-rank features of $\bbZ$ into the model as opposed to merely considering a sparse model of $\bbZ$. Notice, however, that the performance of these two approaches is comparable \cite{oymak2015SimultStruct}. When going from (b) to (c), we see the conspicuous performance improvement entailed by considering the iteratively-reweighted scheme to promote row-sparsity in $\bbZ$. Lastly, when comparing (c) to (d), we gauge the benefits of observing multiple ($P=5$) output signals, especially for large values of $S$ and $L$. In particular, when $S=8$ and $L=5$ we go from a success rate of 0.25 in (c) to a success rate of 0.90 in (d). For this latter setting, exact recovery is achieved consistently for most combinations of $L$ and $S$.

We now consider a weighted undirected graph of the human brain,
consisting of $N=66$ nodes or regions of interest (ROIs) and whose edge weights are given by the density of
anatomical connections between regions~\cite{hagmann2008mapping}. The level of activity of each ROI can be
represented by a graph signal $\bbx$, thus successive
applications of $\bbS$ model a linear evolution of the brain activity pattern. Supposing we observe a
linear combination (filter) of the evolving states of an originally sparse brain signal, then
blind identification amounts to jointly estimating which regions were originally active, the activity in these regions, and the coefficients of the
linear combination. We mimic the recovery rate analysis performed for Erd\H{o}s-R\'{e}nyi graphs; see Fig.~\ref{F:num_exp}(a)-(d) (bottom). As expected, the success rates increase gradually when going from (a) to (d), as we consider more sophisticated algorithms. Furthermore, when comparing the bottom plots in Figs.~\ref{F:num_exp}(a)-(d) with their top counterparts, it is immediate that, for fixed $L$ and $S$, recovery in the brain network is more challenging than in Erd\H{o}s-R\'{e}nyi graphs. This can be explained by the marked structure of the brain network where nodes are divided into two weakly connected hemispheres. Hence, the output signals in one hemisphere are not very informative about the input signals in the opposite hemisphere, {rendering recovery more difficult.}

\vspace{0.1cm}
\noindent\textbf{Comparison with alternative methods.}
So far we have assumed that we observe the entire output signal $\bby$ when trying to infer $\bbx$ and $\bbh$. Nevertheless, it can be the case that we can only sample a subset of the nodes of the graph and try to recover $\{\bbx_0,\bbh_0\}$ from {this reduced number of} observations. Specifically, in Fig.~\ref{F:num_exp}(e) we fix $L\!=\!S\!=\!3$, $P\!=\!1$, and analyze the error behavior (median errors across 50 realizations) as a function of the number of accessible values $y_i$ of the output for different recovery algorithms. Apart from our convex relaxation approach, we consider a naive least squares (LS) baseline where we solve \eqref{E:Filter_input_output_time_khatri} via a pseudoinverse. Moreover, we consider an alternating minimization (AM) algorithm entailing two steps per iteration: i) given $\bbx$, vector $\bbh$ is found as the LS solution of \eqref{E:Filter_input_output_freq}; and ii) given $\bbh$, vector $\bbx$ is found by minimizing $\| \bbx \|_1$ subject to \eqref{E:Filter_input_output_freq} followed by a thresholding operation to retain $S$ non-zero values. These two steps are repeated until convergence, and the algorithm is initialized with the LS estimate of $\bbh$. Finally, we consider as a benchmark our convex method when the support of $\bbx_0$ is known (k.s.)~\cite{SSGMAMAR_camsap15}.

Our proposed method clearly outperforms the naive LS and AM approaches. Say for 60 observations (6 less than the total number of nodes), our method achieves a median error of 0 while the {median} errors for AM and LS are 0.52 and 0.82, respectively. Outperforming LS is not surprising since this algorithm is agnostic to the sparsity in $\bbx_0$ and to the fact that $\bbx_0 \bbh_0^T$ is a rank-one matrix. Further, notice that our method outperforms AM even though the latter assumes that the value of $S$ -- but not the support of $\bbx_0$ -- is known. The big gap between the yellow and the purple curves represents {the performance penalty due to $\text{supp}(\bbx_0)$ being unknown.} For situations where partial information about the support is available, e.g. we know a priori that the input signal is null on a subset of nodes, intermediate curves {(not shown in the figure)} are obtained.

Finally, we include two dashed curves to assess the performance of our approach (with known and unknown support), when the partial observations of $\bby$ are noisy. More specifically, we define $\widetilde{\bby}$ as a perturbed version of $\bby$ given by $\widetilde{\bby} = \bby + \sigma \, \bby \circ \bbr$ where $\sigma$ controls the magnitude of the perturbation and $\bbr$ is a random vector whose entries are drawn independently from a standard normal distribution. In particular, the dashed curves in Fig~\ref{F:num_exp}(e) correspond to $\sigma = 0.01$. As expected, the median reconstruction errors from noisy observations are larger than their noiseless counterparts, however, our approach can be seen to be robust to noise. For intermediate number of observations, the performance of the noisy scheme with unknown support is within $5\%$ of the noiseless one and, once we observe the whole 66 nodes of the output signal, the error of the noisy approach is in the order of $\sigma$ and comparable to that obtained by the noisy scheme with known support.

\begin{figure*}
	\centering
	
	\begin{minipage}[c]{.32\textwidth}
		\includegraphics[width=\textwidth]{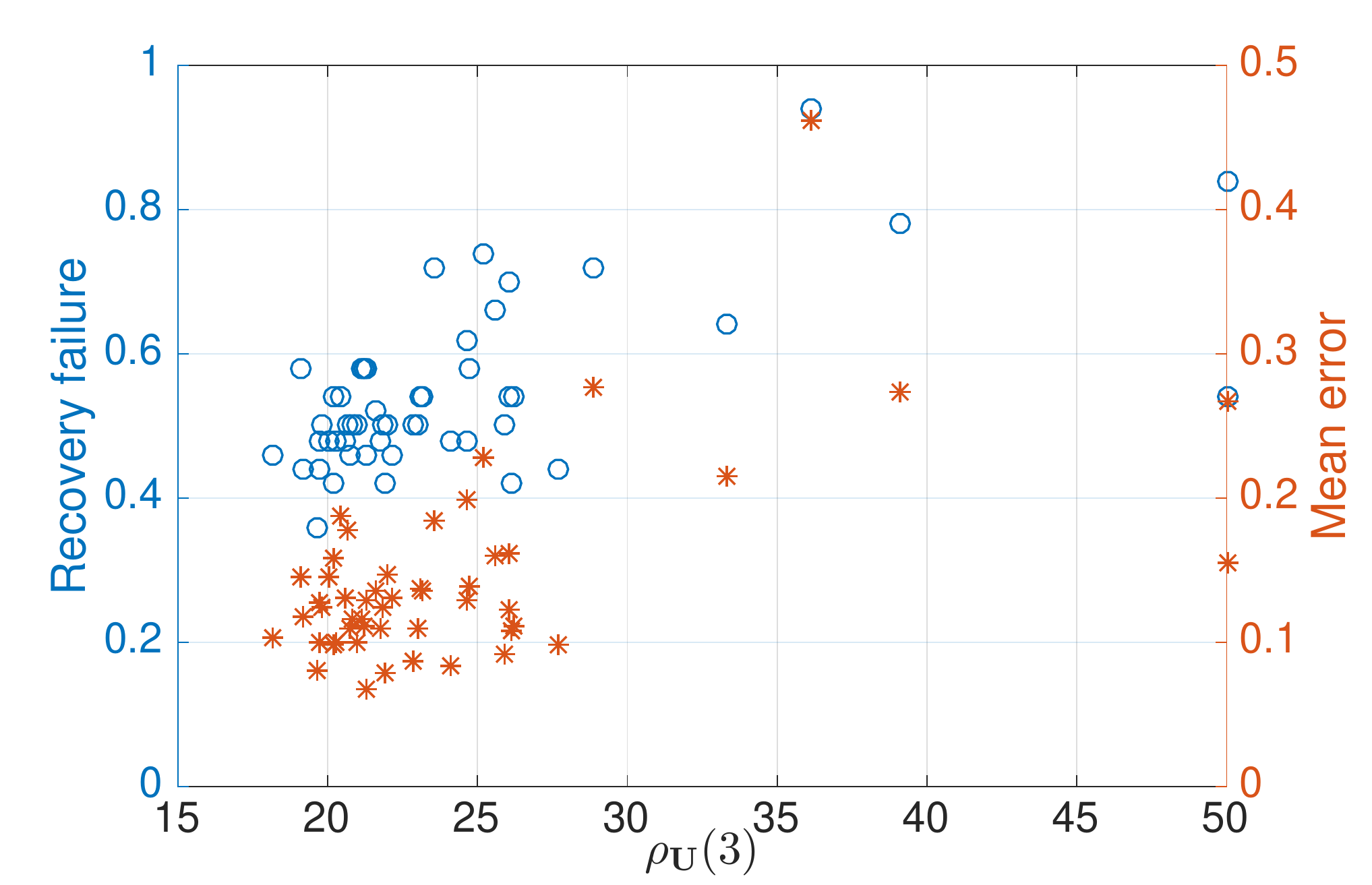}
		\centering{\small (a)}
	\end{minipage}%
	\begin{minipage}[c]{.325\textwidth}
		\includegraphics[width=\textwidth]{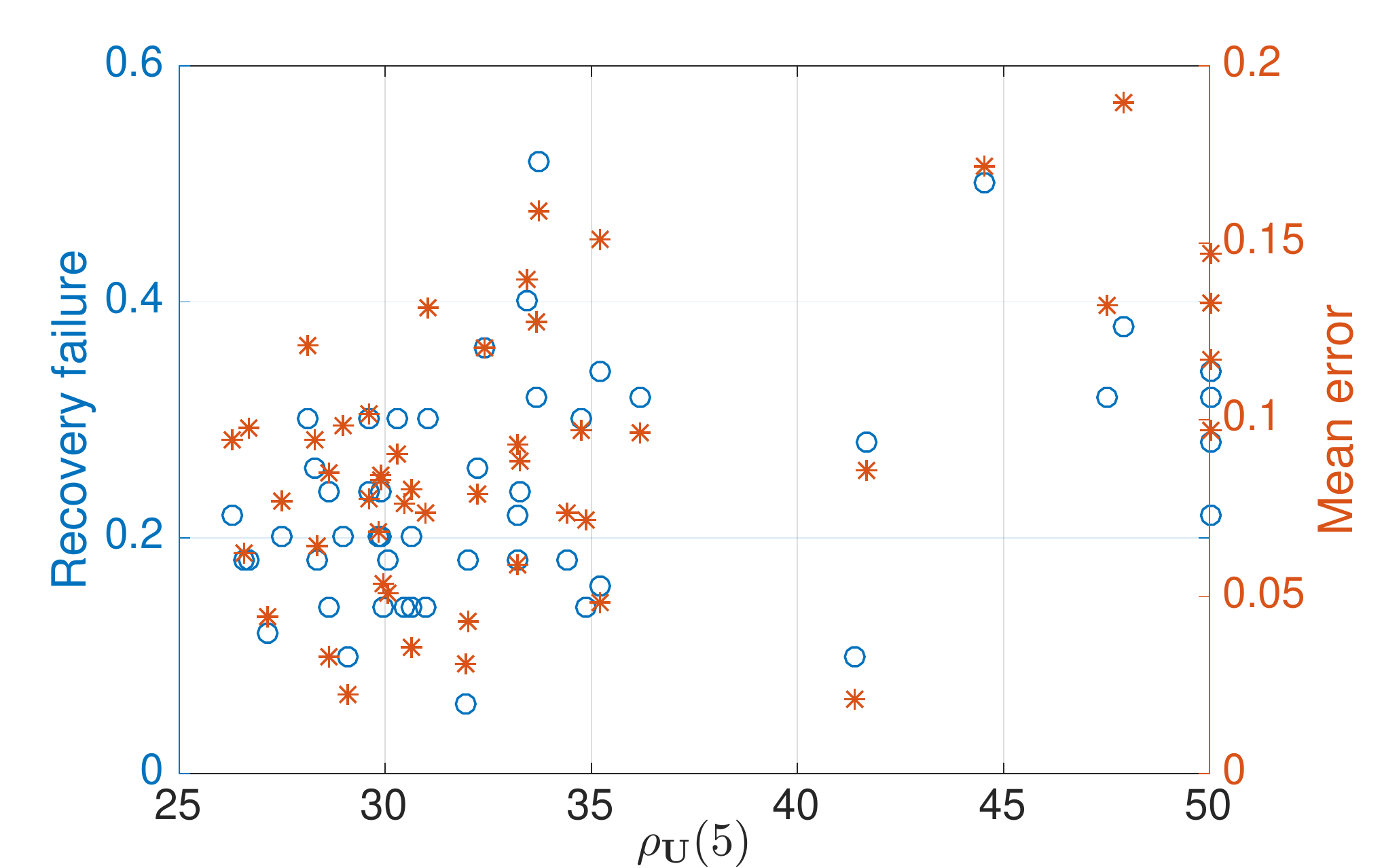}
		\centering{\small (b)}
	\end{minipage}%
	\begin{minipage}{.325\textwidth}
		\includegraphics[width=\textwidth]{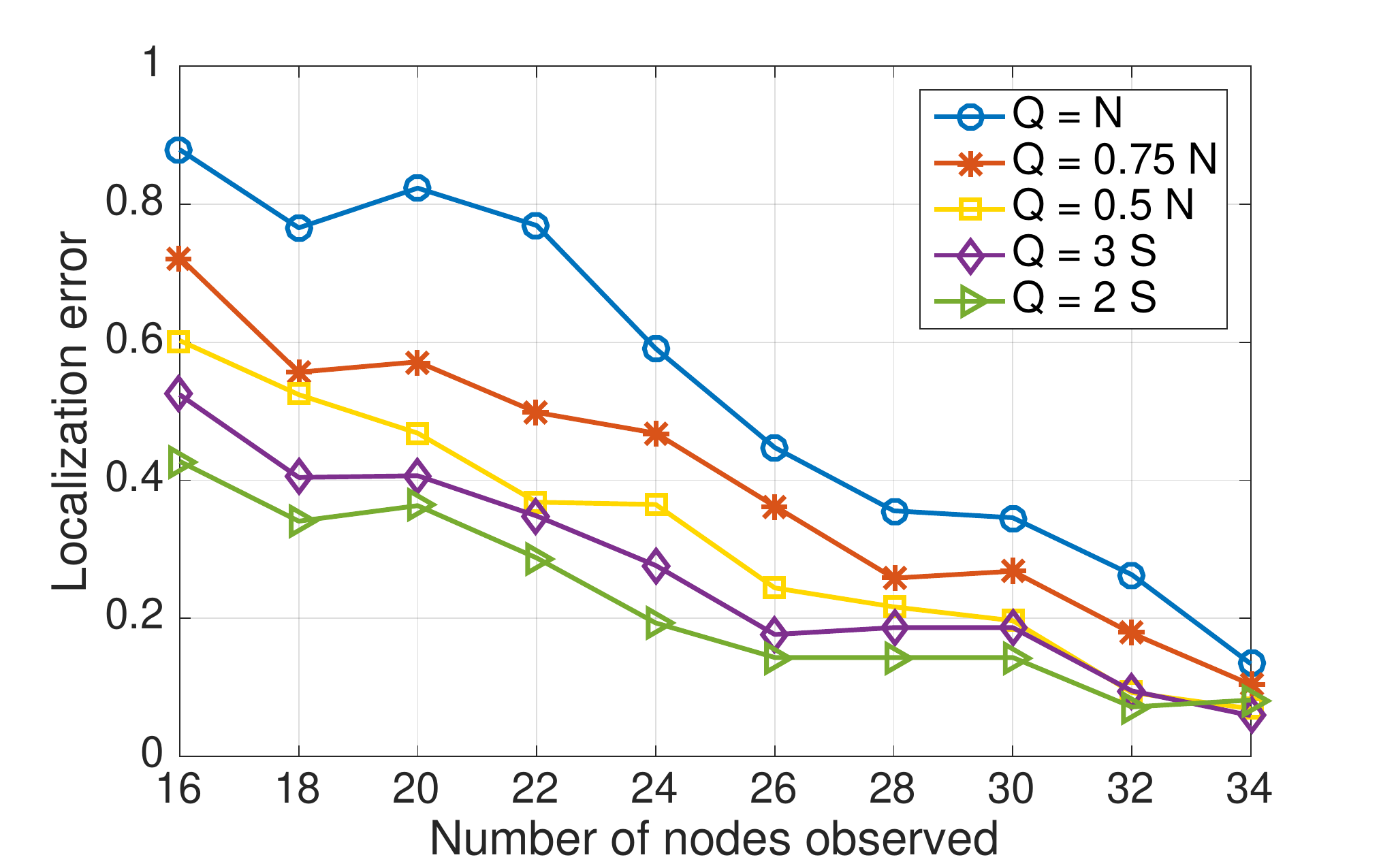}
		\centering{\small (c)}
	\end{minipage}

	\caption{(a) Recovery failure rate and mean reconstruction error of the $\ell_1$ minimization [cf.~\eqref{E:l1_norms_min}] as a function of $\rho_{\bbU}(3)$ for $L=3$ and $S=3$. Each point corresponds to an Erd\H{o}s-R\'enyi random graph of size $N=50$ and $p \in [0.05, 0.15]$. The failure rates and mean errors are computed based on blind identification of 50 graph filters on each graph. (b) Counterpart of (a) for the reweighted $\ell_{2,1}$ plus nuclear norm [cf.~\eqref{E:reweighted_problem}] as a function of $\rho_{\bbU}(5)$ for $L=4$ and $S=5$. (c) Localization error for epidemic sources as a function of the number of observed nodes and parametrized by the number of potential sources~$Q$.}

	\label{F:num_exp_2}
\end{figure*}

\vspace{0.1cm}
\noindent\textbf{Recovery dependence on graph structure.} Theorem~\ref{T:noise_free} reveals that the recovery performance of scheme \eqref{E:l1_norms_min} depends on the graph structure. In particular, when recovering an $S$-sparse input signal, the value of $\rho_{\bbU}(S)$ plays an important role{, with lower values of $\rho_{\bbU}(S)$ leading} to larger values of $\alpha$ and, hence, larger probabilities of recovery [cf.~\eqref{E:main_result_recoverability_alpha_1}  and~\eqref{E:main_result_recoverability_prob}]. Even though Theorem~\ref{T:noise_free} refers to theoretical bounds on the performance, we see that in practice the value of $\rho_{\bbU}(S)$ correlates with the recovery success when implementing \eqref{E:l1_norms_min}; see Fig.~\ref{F:num_exp_2}(a). More specifically, we generate 50 Erd\H{o}s-R\'enyi graphs of size $N=50$ and probability of edge appearance $p$ drawn uniformly from $[0.05, 0.15]$. For each of these graphs we simulate 50 blind-identification problems with $L=3$ and $S=3$, solve them using \eqref{E:l1_norms_min}, and record the reconstruction RMSE and whether the recovery was successful, i.e., RMSE smaller than 0.01. In Fig.~\ref{F:num_exp_2}(a) we plot for each graph the proportion of unsuccessful recoveries (blue circles) and the mean RMSE (orange stars) as a function of $\rho_{\bbU}(3)$. First notice that the values of $\rho_{\bbU}(3)$ are clearly larger than their theoretical lower bound of 3 (cf.~Lemma~\ref{L:behavior_rho} and Fig.~\ref{F:rho}). Most importantly, blind identification is more challenging on graphs with high values of $\rho_{\bbU}(3)$. For example, the average failure rate for the graphs with $\rho_{\bbU}(3) \leq 25$ is $0.51$ whereas for the rest of the graphs the average is $0.64$. Similarly, the average RMSE for the former class of graphs is $0.12$, whereas the graphs with larger values of $\rho_{\bbU}(3)$ achieve an average RMSE of $0.20$.

Although Theorem~\ref{T:noise_free} specifies theoretical bounds of recovery for $\ell_1$ minimization, in practice the performance of more sophisticated schemes such as the reweighted $\ell_{2,1}$ plus nuclear norm minimization also {depends} on the value of $\rho_{\bbU}(S)$; see Fig.~\ref{F:num_exp_2}(b). We mimic the recovery rate analysis for $\ell_1$ minimization but in this case we consider a more challenging blind-identification problem ($L=4$ and $S=5$) so that the failure rates are significant [cf.~Fig.~\ref{F:num_exp_2}(b)]. As can be seen from the figure, the values of $\rho_{\bbU}(S)$ influence the recovery performance. For graphs with values $\rho_{\bbU}(5) \leq 35$ for instance, the average failure rate and average RMSE are $0.22$ and $0.08$, respectively, whereas these values are $0.30$ and $0.12$ for the remaining graphs.

\vspace{0.1cm}
\noindent\textbf{Epidemics.}
We consider the social network of Zachary's karate club \cite{Zachary1977} represented by a graph $\ccalG$ -- with adjacency $\bbA$ -- consisting of 34 nodes or members of the club and 78 undirected edges symbolizing friendships among members. On $\ccalG$ we simulate an $N$-intertwined SIS epidemic model \cite{VanMieghem2011SIS}, which is a susceptible-infected-susceptible (SIS) dynamical process where contagion can only occur among friends (i.e., neighbors) in the social network. To be more precise, each individual at any point in time can be in one of two states: susceptible or infected. At each time point, an infected individual heals with probability $\omega$, thus becoming susceptible. By contrast, a susceptible person $i$ becomes infected with probability $\beta\, |\ccalI_i|$, where $|\ccalI_i|$ is the number of neighbors of $i$ that are infected at that point in time. Our goal is to use the blind graph-filter identification methods to identify the sources of the epidemic outbreak. Even though this model entails non-linear and stochastic dynamics, its mean-field approximation can be estimated by a linear function \cite{VanMieghem2011SIS, Pasqualetti2014Control}. Formally, denoting by $\bbp_t$ a vector collecting the probability of infection of each individual at time $t$, then for small values of $\bbp_t$ we have that 
\begin{equation}\label{E:mf_approx}
\bbp_t \approx [\bbI_N - \upsilon_t ( \omega \bbI_N - \beta \bbA)] \, \bbp_{t-1}
\end{equation} 
where $\upsilon_t$ is a time-varying step-size that arises after discretization of the dynamics. We generally let $\upsilon_t$ vary with time, representing that some time instants could be very active, with multiple contagions and recoveries, whereas in other time instants the epidemic remains more dormant. Equivalently, a time-varying $\upsilon_t$ can be interpreted as having time-varying healing and infection probabilities but maintaining a fixed ratio between them. Defining $\bbS := \omega \bbI_N - \beta \bbA$, we leverage \eqref{E:mf_approx} to write the probabilities of infection at time $T$ as
\begin{equation}
\bbp_T = \prod_{t = 0}^{T-1} (\bbI_N - \upsilon_t \bbS) \bbp_0.
\end{equation}
Notice that the above matrix product is a polynomial in $\bbS$ -- hence a graph filter -- of degree $T$ whose coefficients are a function of $\{\upsilon_t\}_{t=0}^{T-1}$, which we assume to be unknown. Moreover, since we are interested in identifying the original sources of infection, the initial probability vector $\bbp_0$ is what we want to estimate. If we were to observe $\bbp_T$, then the problem is precisely an instance of blind graph-filter identification. Given that observing the probabilities of infection at time $T$ seems impractical, we consider a case where we observe $W$ epidemic outbreaks in the same population and estimate $\bbp_T$ from these. Notice that the epidemic outbreaks need not refer to diseases but could model the spread of rumors or the adoption of new technologies. We want to emphasize that the number and identities of the sources in each of these outbreaks are generally different, but what remains constant is the probability $\bbp_0$ with which the sources are chosen. Nevertheless, we assume that the nodes that \emph{could} start an epidemic are just a subset of the total (of cardinality $S$), so that $\bbp_0$ is sparse. 
For the simulations here, we consider $T=3$, $W=500$, and $S$ drawn at random from $\{3, 4, 5\}$. After estimating $\bbp_T$ from the realizations, we implement our reweighted $\ell_{2,1}$ plus nuclear norm minimization [cf.~\eqref{E:reweighted_problem}] to obtain $\tilde{\bbp}_0$, our estimate of $\bbp_0$, which we use to identify the potential sources of contagion. In particular, we quantify the localization error as $\| \mathrm{supp}(\tilde{\bbp}_0) - \mathrm{supp}({\bbp_0}) \|_0 / S$, which specifies the proportion of misidentified sources. 
Assuming that in each realization we can only access the state (infected or susceptible) of a subset of nodes, in Fig.~\ref{F:num_exp_2}(c) we plot the localization error -- mean across 20 realizations -- as a function of the number of nodes observed.
In addition, we consider the incorporation of different levels of a priori information. Specifically, we analyze scenarios where we know that the $S$ sources belong to a subset of potential nodes of cardinality $Q$, so that when $Q = N$ no a priori information is considered. 

Fig.~\ref{F:num_exp_2}(c) shows that, independently of the value of $Q$, the localization error decreases when the number of observed nodes increases. Intuitively, the larger the number of observed nodes, the more complete the estimation of $\bbp_T$ is, hence, entailing a more reliable estimate $\tilde{\bbp}_0$ of the epidemic sources. For $Q=N$, if only $16$ nodes are observed the localization is very poor with an average error of $0.88$, but when all $34$ nodes are observed this error decreases to $0.13$. Furthermore, Fig.~\ref{F:num_exp_2}(c) illustrates the benefit of a priori information. For instance, when only $24$ nodes are observable and we have no a priori information, the average error is $0.59$. However, this error can be reduced to $0.37$ and $0.19$ by constraining the potential sources to subsets of cardinality $Q = N/2$ and $Q = 2 S$, respectively.

\section{Conclusions}\label{S:Conclusions}

We formulated and studied the problem of blind graph-filter identification, an extension of blind deconvolution of time (or spatial) domain signals to graphs. Envisioned application domains encompass a particular class of bilinear inverse problems, where the observed graph signal is modeled as a linear combination of  diffusion processes driven by a limited number of sources. While the said observations are bilinear functions of the filter coefficients and the sparse input signal, leveraging the frequency interpretation of graph signals and graph filters it was possible to show that they are also linearly related to the entries of a lifted rank-one, row-sparse matrix of the unknowns. Accordingly, the blind graph-filter
identification problem was tackled via rank and sparsity minimization subject to linear constraints, an 
inverse problem amenable to convex relaxations offering provable recovery guarantees under simplifying assumptions.  Numerical tests validated the theoretical claims and demonstrated that the proposed approach offers satisfactory recovery performance, even for settings well beyond the scope of the analysis.  Future research includes also estimating
the shift operator by bringing to bear methods of network topology inference~\cite{kolaczyk2009book};
see also~\cite{meigraphstructure,segarra2016networktopology} for recent related approaches to estimate the graph structure from graph signals.

\begin{appendices}

\section{Proof of Theorem~\ref{T:noise_free}}\label{App:proof_the_noise_free}

We reiterate that the proof in this section follows closely the ideas used to establish~\cite[Theorem 3.1]{LingBiConvexCS}; see also~\cite{ahmed2014BlindDeconvConvex}. Thus the emphasis will be on the differences that arise in the
graph setting dealt with here.

\noindent\textbf{Preliminary definitions and notations}. Going back to the discussion in Section \ref{S:BlindIdent}, for $\bbZ=\bbx\bbh^T$ recall the linear mapping $\ccalM:\mathbb{C}^{N\times L}\mapsto \mathbb{C}^N$ such that
\begin{equation}\label{E:linear_map}
	\widehat{\bby}=\ccalM(\bbZ):=\{\bbu_i^T
	\bbZ\bbpsi_i\}_{i=1}^N,
\end{equation}
where $\bbu_i^T$ and $\bbpsi_i^T$ denote the $i$-th rows of $\bbU$ and $\bbPsi$, respectively. Whenever clear from context, as in \eqref{E:linear_map}, notation $\{ \cdot \}$ will be used for the element-wise definition of a vector.  Accordingly, one can write 
$\widehat{\bby}=\bbM\text{vec}(\bbZ)$ for $\bbM:=\big(\bbPsi^T\odot\bbU^T\big)^T\in \mathbb{C}^{N\times LN}$, where
\begin{equation}\label{E:M_Her_cols}
	\bbM^H=[\bbm_1,\ldots,\bbm_N]\in\mathbb{C}^{LN\times N},\quad \bbm_i= \overline{\bbpsi}_i\otimes\overline{\bbu}_i.
\end{equation}
By using the inner product defined on $\mathbb{C}^{N\times L}$ as $\langle\bbX,\bbZ\rangle:=\text{tr}(\bbX\bbZ^H)$, the adjoint
operator $\ccalM^*:\mathbb{C}^N\mapsto\mathbb{C}^{N\times L}$ as well as $\ccalM^*\ccalM:\mathbb{C}^{N\times L}\mapsto\mathbb{C}^{N\times L}$ take the form
\begin{equation}\label{E:adjoint_composition}
	\ccalM^*(\bbz)=\sum_{i=1}^Nz_i\overline{\bbu}_i\bbpsi_i^H,\quad \ccalM^*\ccalM(\bbZ)=\sum_{i=1}^N\overline{\bbu}_i\bbu_i^T\bbZ\bbpsi_i\bbpsi_i^H.
\end{equation}
Based on the aforementioned definitions, the following useful formula holds
\begin{align}\label{E:vectorized_compostion}
	\text{vec}(\ccalM^*\ccalM(\bbZ))={}&\bbM^H\bbM\text{vec}(\bbZ)\nonumber\\
	={}&\sum_{i=1}^N\left(\overline{\bbpsi}_i\bbpsi_i^T\otimes\overline{\bbu}_i
	\bbu_i^T\right)\text{vec}(\bbZ).
\end{align}

Consider a subset of measurements $\Gamma_p\subset\{1,\ldots,N\}$ with $|\Gamma_p|=Q$, and define the row-sampled operator $\ccalM_p:\mathbb{C}^{N\times L}\mapsto \mathbb{C}^Q$ as $\ccalM_p(\bbZ):=\{\bbu_i^T\bbZ\bbpsi_i\}_{i\in\Gamma_p}$;
as well as $\ccalM_p^*\ccalM_p(\bbZ):=
\sum_{i\in\Gamma_p}\overline{\bbu}_i\bbu_i^T\bbZ\bbpsi_i\bbpsi_i^H$. The linear operator $\ccalM_p$ has matrix representation $\bbM_p\in\mathbb{C}^{Q\times LN}$, where $\bbM_p^H$ has columns $\{\bbm_i\}_{i\in\Gamma_p}$. Accordingly, one can write $\ccalM_p(\bbZ)=\bbM_p\text{vec}(\bbZ)$ and $\text{vec}(\ccalM_p^*\ccalM_p(\bbZ))=\bbM_p^H\bbM_p\text{vec}(\bbZ)$.

Regarding the support of the sparse vector $\bbx_0$, we can assume without loss of generality that the first $S$ entries of $\bbx_0$ are non-zero and so are
the first $S$ rows of $\bbZ_0=\bbx_0\bbh_0^T$. While slightly abusing notation, it is convenient to denote as $\Omega$ the supports of both
$\bbx_0$ and $\bbZ_0$. Next, let $\bbx_\Omega$ and $\bbZ_\Omega$ denote the orthogonal projections of $\bbx$ and $\bbZ$ onto $\Omega$, respectively. Leveraging these definitions, we can define $\ccalM_\Omega$ and $\ccalM_\Omega^*$ as the restriction of $\ccalM$ and $\ccalM^*$ to $\Omega$, such that
\begin{align}
	\ccalM_\Omega(\bbZ)={}&\{\bbu_i^T\bbZ_\Omega\bbpsi_i\}_{i=1}^N=\{\bbu_{i,\Omega}^T\bbZ_\Omega\bbpsi_i\}_{i=1}^N,\nonumber\\
	\ccalM_\Omega^*(\bbz)={}&\sum_{i=1}^Nz_i\overline{\bbu}_{i,\Omega}\bbpsi_i^H.\label{E:restrictions_omega}
\end{align}
Naturally, one can be interested in projected \emph{and} row-sampled operators of the form $\ccalM_{p, \Omega} =\{\bbu_{i,\Omega}^T\bbZ_\Omega\bbpsi_i\}_{i \in \Gamma_p}$.

Instrumental to our proof will be the following form of the non-commutative Bernstein inequality for matrices~\cite{TroppMatrixConcentration}.

\begin{mytheorem}[Th. 4.4~\cite{LingBiConvexCS}]\label{T:bernstein} Consider a finite sequence $\{\bbZ_i\}_{i\in\Gamma_p}$ of independent, centered random matrices with dimension $M\times M$. Assume that $\|\bbZ_i\|\leq R$ and introduce the random matrix
	\begin{equation*}
		\bbSigma=\sum_{i\in\Gamma_p}\bbZ_i
	\end{equation*}
	with variance parameter
	\begin{equation}\label{E:def_sigma}
		\sigma^2=\max\Big\{\Big\|\sum_{i\in\Gamma_p}\E{\bbZ_i\bbZ_i^H}\Big\|,\Big\|\sum_{i\in\Gamma_p}\E{\bbZ_i^H\bbZ_i}\Big\|\Big\}.
	\end{equation}
	Then for all $t\geq 0${, it holds that}
	\begin{equation}
		\Pc{\|\bbSigma\|\geq t}\leq 2M\exp\left(-\frac{t^2/2}{\sigma^2+Rt/3}\right).
	\end{equation}
\end{mytheorem}
\vspace{1mm}
\noindent\textbf{Optimality conditions.} A key step in the proof is Proposition~\ref{P:DetermRecov_L1_4conditions}, which states four conditions guaranteeing that the unique solution to \eqref{E:l1_norms_min} is indeed $\bbZ_0=\bbx_0\bbh_0^T$. See \cite[Proposition 4.2]{LingBiConvexCS} for a proof. In what follows, we denote by $\Omega^\perp$ the orthogonal complement of $\Omega$.

\begin{myproposition}\label{P:DetermRecov_L1_4conditions}
	The rank-one matrix $\bbZ_0 = \bbx_0\bbh_0^T$ is the unique minimizer of \eqref{E:l1_norms_min}, if there exists $\bbY \in \mathrm{range}(\ccalM^*)$ and a scalar $\gamma$ that satisfy conditions:\\ a) $\|\sign(\bbZ_0) -\bbY_{\Omega}\|_\mathrm{F} \leq 1/(4\sqrt{2}\gamma)$,         b) $\|\bbY_{\Omega^\perp}\|_\infty \leq 1/2$; \\
	and the operator $\ccalM$ satisfies conditions:\\
		c) $\|\ccalM_{\Omega}^*\ccalM_{\Omega}-\bbI_{\Omega}\| \leq 1/2$, d) $\|\ccalM \|\leq \gamma$.
\end{myproposition}

The rest of the section is devoted to {establish} each of the four conditions \emph{a)}-\emph{d)} given above. Although Proposition \ref{P:DetermRecov_L1_4conditions} entails deterministic conditions, as customary when dealing with sparse recovery algorithms we show that conditions \emph{a)}-\emph{d)} can be satisfied with a certain probability. More specifically, it turns out that all conditions hold with probability at least $1 - N^{-\alpha + 1}$, giving rise to the statement in Theorem~\ref{T:noise_free} [cf.~\eqref{E:main_result_recoverability_prob}].

We begin by showing condition \emph{c)} in Proposition~\ref{P:DetermRecov_L1_4conditions}. We do so by first showing a more general result in Lemma~\ref{lemma_4_7}, which is a modification of \cite[Lemma 4.7]{LingBiConvexCS}.

\begin{mylemma}\label{lemma_4_7}
For any fixed $0 < \delta \leq 1$ and partition $\{ \Gamma_p \}_{p=1}^P$ of $\{ 1, 2, \ldots, N\}$ with $| \Gamma_p | = Q$ for all $p$, and defining $\bbT_p = \sum_{i \in \Gamma_p} \overline{\bbpsi}_i \bbpsi_i^T$ we have that
\begin{equation}\label{E:statement_lemma_4_7}
\max_{1\leq p \leq P} \sup_{\| \bbZ_\Omega \|_\mathrm{F} = 1}  \| \ccalM^*_{p, \Omega} \ccalM_{p, \Omega} (\bbZ) - \bbZ_\Omega \bbT_p \|_\mathrm{F} \leq \frac{\delta Q}{N},
\end{equation}
with probability at least $1 - N^{-\alpha + 1}$ if $\alpha \geq 1$ and 
\begin{equation}\label{E:statement_lemma_4_7_alpha}
\alpha \leq \frac{Q \delta^2 (5/2 + 2\delta/3)^{-1}} {\rho_{\bbPsi}(L) \rho_{\bbU}(S) N \log(2NLS)}.
\end{equation}
\end{mylemma}
\begin{myproof}
Define $\bbUpsilon_i \in \mathbb{C}^{LN \times LN}$ as
\begin{equation}\label{E:def_eta_i}
\bbUpsilon_i:= (\overline{\bbpsi}_i \bbpsi_i^T) \otimes (\overline{\bbu}_{i,\Omega} \bbu_{i,\Omega}^T - \bbI_{N,\Omega}),
\end{equation}
and notice that every $\bbUpsilon_i$ is a centered random matrix since $\E{\overline{\bbu}_{i,\Omega} \bbu_{i,\Omega}^T} = \bbI_{N,\Omega}$ due to the normalization of $\bbU$ (cf. Theorem~\ref{T:noise_free}). Leveraging \eqref{E:vectorized_compostion}, it follows that
%
$\mathrm{vec}(\ccalM^*_{p, \Omega} \ccalM_{p, \Omega} (\bbZ) - \bbZ_\Omega \bbT_p) = \sum_{i \in \Gamma_p} \bbUpsilon_i \mathrm{vec}(\bbZ_\Omega)$ and, from the definition of matrix induced norm, we obtain that
%
\begin{equation}\label{E:lemma_4_7_proof_040}
\sup_{\| \bbZ_\Omega \|_\mathrm{F} = 1}  \| \ccalM^*_{p, \Omega} \ccalM_{p, \Omega} (\bbZ) - \bbZ_\Omega \bbT_p \|_\mathrm{F} = \Big\| \sum_{i \in \Gamma_p} \bbUpsilon_i \Big\|.
\end{equation}
Thus, our goal is to use Theorem~\ref{T:bernstein} to find a probabilistic bound on $\| \sum_{i \in \Gamma_p} \bbUpsilon_i \|$. To this end, we need to find suitable expressions for $R$ and $\sigma^2$. For computing $R$, notice that
\begin{equation}\label{E:lemma_4_7_proof_050}
\| \bbUpsilon_i \| = | \bbpsi_i^T \overline{\bbpsi}_i | \| \overline{\bbu}_{i,\Omega} \bbu_{i,\Omega}^T - \bbI_{N,\Omega} \| \leq \rho_{\bbPsi}(L) \rho_{\bbU}(S),
\end{equation}
where we used the definition of {$\rho_\bbA$} in \eqref{E:def_rho}. Since \eqref{E:lemma_4_7_proof_050} is true for all $i$, we have that $R \leq \rho_{\bbPsi}(L) \rho_{\bbU}(S)$ as wanted. In order to compute $\sigma^2$, it suffices to consider $\bbUpsilon_i \bbUpsilon_i^H$ since $\bbUpsilon_i$ is a hermitian matrix for all $i$ [cf.~\eqref{E:def_sigma}]. Thus, we have
\begin{align}\label{E:lemma_4_7_proof_060}
&\sigma^2 = \Big\| \sum_{i\in\Gamma_p}\E{\bbUpsilon_i\bbUpsilon_i^H} \Big\|  \\
&\leq \rho_{\bbPsi}(L) \Big\| \sum_{i\in\Gamma_p}   (\overline{\bbpsi}_i \bbpsi_i^T) \otimes \E{(\overline{\bbu}_{i,\Omega} \bbu_{i,\Omega}^T - \bbI_{N,\Omega})^2} \Big\| \nonumber.
\end{align}
To compute the required expected value, expand the square and use the facts that $\bbu_{i,\Omega}^T \overline{\bbu}_{i,\Omega} \leq \rho_{\bbU} (S)$ and $\E{\overline{\bbu}_{i,\Omega} \bbu_{i,\Omega}^T} = \bbI_{N, \Omega}$. Substituting these into \eqref{E:lemma_4_7_proof_060} and recalling the definition of $\bbT_p$ from {the statement of Lemma \ref{lemma_4_7},} it follows that
\begin{equation}\label{E:lemma_4_7_proof_065}
\sigma^2 \leq \rho_{\bbPsi}(L) \rho_{\bbU}(S) \| \bbT_p \| \leq \rho_{\bbPsi}(L) \rho_{\bbU}(S) \frac{5Q}{4N},
\end{equation}
where the last inequality follows from \cite[equation (4.7)]{LingBiConvexCS}. We use the results in \eqref{E:lemma_4_7_proof_050} and \eqref{E:lemma_4_7_proof_065} to apply Theorem~\ref{T:bernstein} for the case where $t = \delta Q/N$. In using Theorem~\ref{T:bernstein}, $\bbUpsilon_i$ can be interpreted as an $LS \times LS$ matrix for all $i$ since this is the dimension of its support [cf.~\eqref{E:def_eta_i}]. Thus, we obtain
\begin{align}\label{E:lemma_4_7_proof_070}
\Pc{\Big\| \! \sum_{i\in\Gamma_p} \bbUpsilon_i \Big\| \geq \frac{\delta Q}{N} \!} \! \leq \! 2 L S \exp \left(  - \frac{\delta^2 Q/2N}{\rho_{\bbPsi}(L) \rho_{\bbU}(S) (\frac{5}{4}+\frac{\delta}{3})} \right) \! .
\end{align}
If, in particular, for some constant $\alpha \geq 1$ we choose 
\begin{equation}\label{E:lemma_4_7_proof_080}
Q \geq \frac{\alpha}{\delta^2} \left( \frac{5}{2} + \frac{2 \delta}{3} \right) \rho_{\bbPsi}(L) \rho_{\bbU}(S) N \log(2NLS),
\end{equation}
the right-hand side in \eqref{E:lemma_4_7_proof_070} is not larger than $N^{-\alpha}$. Using a union bound on \eqref{E:lemma_4_7_proof_070} to find the probability of $\| \sum_{i\in\Gamma_p} \bbUpsilon_i \| \geq \delta Q/L$ for \emph{some} $p$, and then considering its complementary event, we get that
\begin{align}\label{E:lemma_4_7_proof_090}
\Pc{\max_{1 \leq p \leq P} \Big\| \! \sum_{i\in\Gamma_p} \bbUpsilon_i \Big\| \leq \frac{\delta Q}{N} \!} \geq 1 - P N^{-\alpha} \geq 1 - N^{-\alpha + 1}.
\end{align}
From \eqref{E:lemma_4_7_proof_040} and \eqref{E:lemma_4_7_proof_090}, statement \eqref{E:statement_lemma_4_7} follows whereas the expression for $\alpha$ in \eqref{E:statement_lemma_4_7_alpha} is obtained from \eqref{E:lemma_4_7_proof_080}.
\end{myproof}

Condition \emph{c)} in Proposition~\ref{P:DetermRecov_L1_4conditions} follows by specializing Lemma~\ref{lemma_4_7} for $\delta = 1/2$ and $P=1$. Notice that the latter equality implies that $Q=N$ and $\bbT_1 = \bbI_L$. As discussed after the statement of Theorem~\ref{T:noise_free}, a novel component introduced in the proof of Lemma~\ref{lemma_4_7} compared to that of \cite[Lemma 4.7]{LingBiConvexCS} is the appearance of $\rho_{\bbU}(S)$ in the lower bound for $Q$ [cf.~\eqref{E:lemma_4_7_proof_080}]. Notice that in \cite{LingBiConvexCS}, $\rho_{\bbU}(S) = S$ since $\bbU$ is assumed to be random Fourier, thus, every element has unit magnitude. In our case, different graphs give rise to less favorable (larger) bounds on $Q$ since $\rho_{\bbU}(S) \geq S$. 

Our next step is to prove condition \emph{d)} in Proposition~\ref{P:DetermRecov_L1_4conditions}. We attain this in the following lemma, a restatement of \cite[Lemma 4.9]{LingBiConvexCS}.

\begin{mylemma}\label{lemma_4_9}
For $\ccalM$ defined in \eqref{E:linear_map} and $\alpha \geq 1$ then it holds that
\begin{equation}\label{E:statement_lemma_4_9}
\| \ccalM \| \leq \gamma := \sqrt{2N(\log(2LN)+1)+1},
\end{equation}
with probability at least $1-N^{-\alpha}$ if $\alpha \leq ( \rho_{\bbPsi}(L) \log(N))^{-1}$.
\end{mylemma}
\begin{myproof}
The proof follows the same steps as those in the proof of \cite[Lemma 4.9]{LingBiConvexCS}. The only step that needs to be checked is whether $\E{(\overline{\bbu}_i \bbu_i^T - \bbI_N)^2} = (N-1) \bbI_N$ still holds in our context. To see this, expand the square and use the facts that $\E{\overline{\bbu}_i \bbu_i^T} = \bbI_N$ and that $\bbu_i^T\overline{\bbu}_i = N$ for all $i$, which are immediate implications of the normalization of $\bbU$ (cf. Theorem~\ref{T:noise_free}). Finally, to obtain the expressions presented in the statement of the proposition notice that $k$, $L$ and $\mu^2_{\max} k /L$ in \cite{LingBiConvexCS} are equal to $L$ and $N$ and $\rho_{\bbPsi}(L)$ here, respectively.
\end{myproof}

\vspace{1mm}
\noindent\textbf{Construction of dual certificate.} 
We construct the inexact dual certificate $\bbY$ mentioned in conditions \emph{a)} and \emph{b)} of Proposition~\ref{P:DetermRecov_L1_4conditions} via the celebrated golfing scheme \cite{Gross2011Golfing}. The goal of the scheme is to generate a sequence of random matrices $\bbY_p$ in $\mathrm{range}(\ccalM^*)$ such that the sequence converges to $\sign(\bbZ_0)$ [cf. \emph{a)}] while keeping each entry in $\Omega^\perp$ small [cf. \emph{b)}]. We initialize $\bbY_0 = \mathbf{0}$ and set the following recursion
\begin{equation}\label{E:recursive_formula_Y}
\bbY_p := \bbY_{p-1} + \frac{N}{Q} \ccalM^*_p \ccalM_p ( \mathrm{sign}(\bbZ_0) - \bbY_{p-1,\Omega})
\end{equation}
where the successive operators $\ccalM_p$ are based on different blocks $\Gamma_p$ of a partition $\{ \Gamma_p \}_{p=1}^P$ of $\{1, 2, \ldots, N \}$ such that $| \Gamma_p | = Q$ for all $p$.
We define our desired dual certificate as $\bbY := \bbY_P$ and the sequence of residuals $\bbW_p := \bbY_{p, \Omega} - \mathrm{sign}(\bbZ_0)$. From this definition, it follows that $\bbW_0 = \bbY_0 - \mathrm{sign}(\bbZ_0) =  - \mathrm{sign}(\bbZ_0)$ implying that $\| \bbW_0 \|_F = \sqrt{LS}$ from the sparsity level in $\bbZ_0$. Furthermore, Lemma~\ref{lemma_4_7} can be leveraged to show that $\| \bbW_p \|_\mathrm{F} \leq 2^{-1} \, \| \bbW_{p-1} \|_\mathrm{F}$ with probability at least $1-N^{-\alpha+1}$ as long as \eqref{E:statement_lemma_4_7_alpha} is satisfied for $\delta = 1/4$ (cf. Lemma 4.6 in \cite{LingBiConvexCS}). Combining these two observations, we obtain that $\| \bbW_P \|_\mathrm{F} = \| \bbY_\Omega - \mathrm{sign}(\bbZ_0) \|_\mathrm{F} \leq 2^{-P} \sqrt{LS}$. By equating this upper bound with that in condition \emph{a)} of Proposition~\ref{P:DetermRecov_L1_4conditions}, we obtain a lower bound on the values of $P$ that guarantee fulfillment of the condition, namely
\begin{equation}\label{E:lower_bound_P}
P \geq \frac{\log(4 \gamma \sqrt{2LS})}{\log(2)}.
\end{equation}
Finally, in order to show condition \emph{b)} in Proposition~\ref{P:DetermRecov_L1_4conditions} we begin by leveraging \eqref{E:recursive_formula_Y} and the definition of $\bbW_p$ to write
\begin{equation}\label{E:rewritten_formula_Y_W}
\bbY_P = \bbY = - \frac{N}{Q} \sum_{p=1}^P \ccalM^*_p \ccalM_p( \bbW_{p-1}).
\end{equation}
In order to show that $\| \bbY_{\Omega^\perp} \|_\infty \leq 1/2$, it suffices to show that $ \| [\ccalM^*_p \ccalM_p( \bbW_{p-1})]_{\Omega^\perp} \|_\infty \leq 2^{-p-1} Q/N$ and then use the expression of a geometric sum in \eqref{E:rewritten_formula_Y_W}. We show this latter claim via the following lemma, which is a modified version of \cite[Theorem 4.11]{LingBiConvexCS}.

\begin{mylemma}\label{L:Theorem_4_11}
Under the assumption that $\| \bbW_p \|_\mathrm{F} \leq 2^{-p} \sqrt{LS}$, it holds that
\begin{equation}\label{E:statement_theorem_4_11}
\Pc{ \! \| [\ccalM^*_p \ccalM_p( \bbW_{p-1})]_{\Omega^\perp} \|_\infty \leq \frac{Q}{2^{p+1}N} \!} \geq 1 - N^{-\alpha+1}
\end{equation}
for all $p$ if $\alpha \geq 1$ and 
\begin{equation}\label{E:statement_theorem_4_11_alpha}
\alpha \leq \frac{3Q\left( 120 \frac{\rho_{\bbU}(1)\rho_{\bbPsi}(1) L S}{\rho_{\bbU}(S)\rho_{\bbPsi}(L)} + 8 \sqrt{\frac{\rho_{\bbU}(1)\rho_{\bbPsi}(1) L S}{\rho_{\bbU}(S)\rho_{\bbPsi}(L)}}\right)^{-1}}{N\rho_{\bbU}(S)\rho_{\bbPsi}(L) \log(2SN^2)}.
\end{equation}
\end{mylemma}
\begin{myproof}
The proof follows similar steps as those in the proof of \cite[Theorem~4.11]{LingBiConvexCS}. However, in our case, matrix $\bbU$ is not random Fourier, thus, different bounds must be used. In particular, we leverage the definition of {$\rho_\bbA$} in \eqref{E:def_rho} to state that [cf.~\eqref{E:M_Her_cols}] $\| \bbm_i \|_\infty \leq \| {\bbpsi}_i \|_\infty \| {\bbu}_i \|_\infty = \sqrt{\rho_{\bbPsi}(1) \rho_{\bbU}(1)}$ and, similarly, that $\| \bbm_i \|_2 \leq \sqrt{\rho_{\bbPsi}(L) \rho_{\bbU}(S)}$. These bounds lead to the following expressions for $R$ and $\sigma^2$
\begin{align}\label{E:bounds_R_and_sigma_2}
R &\leq 2^{-p+1} \sqrt{\rho_{\bbU}(1) \rho_{\bbPsi}(1) \rho_{\bbU}(S) \rho_{\bbPsi}(L) L S}, \\
\sigma^2 &\leq 2^{-2p} \rho_{\bbU}(1) \rho_{\bbPsi}(1) 5QLS / N.
\end{align}
After this, we apply the Bernstein inequality (cf. Theorem~\ref{T:bernstein}) and implement a union bound mimicking the procedure in \eqref{E:lemma_4_7_proof_070}-\eqref{E:lemma_4_7_proof_090} to attain the statement of the lemma.
\end{myproof}

Thus far, we have found requirements on $\alpha \geq 1$ that ensure the fulfillment of conditions \emph{a)}-\emph{d)} in Proposition~\ref{P:DetermRecov_L1_4conditions} with high probability. Consequently, we need to satisfy the most restrictive of the requirements on $\alpha$ to guarantee simultaneous fulfillment of conditions \emph{a)}-\emph{d)} and, hence, ensure recovery of $\bbZ_0 = \bbx_0 \bbh_0^T$. These requirements are re-stated below, in order of appearance:
\vspace{0.1cm}
\begin{subnumcases}{\!\!\!\!\!\!\!\!\!\!\!\!\!\!\! \alpha \! \leq \!\!\!}
   \!\! \frac{3} {34 \, {\rho_{\bbU}(S)\rho_{\bbPsi}(L)}  \log(2NLS)}, \label{E:cond_alpha_1}
   \\
   \!\!\!\frac{1}{\rho_{\bbPsi}(L) \log(N)}, \label{E:cond_alpha_2} \\
   \!\!\!\!\frac{3 \log(2)/128} {{\rho_{\bbU}(S) \rho_{\bbPsi}(L) }\log(4 \gamma \sqrt{2LS}) \log(2NLS)} \! , \label{E:cond_alpha_3} \\
   \!\!\! \frac{3 \log(2) \! \left( 120 \frac{\rho_{\bbU}(1)\rho_{\bbPsi}(1) L S}{\rho_{\bbU}(S)\rho_{\bbPsi}(L)} + 8 \sqrt{\frac{\rho_{\bbU}(1)\rho_{\bbPsi}(1) L S}{\rho_{\bbU}(S)\rho_{\bbPsi}(L)}}\right)^{\!\!-1}}{{ \rho_{\bbU}(S)\rho_{\bbPsi}(L)} \log(4 \gamma \sqrt{2LS}) \! \log(2N^2S)} \!.\label{E:cond_alpha_4}
\end{subnumcases}
\vspace{0.1cm}

\noindent Recall that \eqref{E:cond_alpha_1} is obtained by specializing \eqref{E:statement_lemma_4_7_alpha} to $Q=N$ and $\delta = 1/2$ [cond.~\emph{c)}] whereas \eqref{E:cond_alpha_2} comes from Lemma~\ref{lemma_4_9} [cond.~\emph{d)}]. Expression \eqref{E:cond_alpha_3} is obtained by particularizing \eqref{E:statement_lemma_4_7_alpha} to $\delta = 1/4$, noting that $Q = N/P$ and using \eqref{E:lower_bound_P} to bound $P$ [cond.~\emph{a)}]. Finally, \eqref{E:cond_alpha_4} is derived from \eqref{E:statement_theorem_4_11_alpha} by again substituting $Q = N/P$ [cond.~\emph{b)}].

Notice that fulfillment of \eqref{E:cond_alpha_3} immediately implies fulfillment of \eqref{E:cond_alpha_1} and \eqref{E:cond_alpha_2}. By leveraging the facts that $N \geq L$, $\rho_{\bbU}(1) S \geq \rho_{\bbU}(S)$, and $\rho_{\bbPsi}(1) L \geq \rho_{\bbPsi}(L)$ (cf.~Lemma~\ref{L:behavior_rho}), it follows that \eqref{E:cond_alpha_4} implies \eqref{E:cond_alpha_3}. This makes \eqref{E:cond_alpha_4} the most stringent of the requirements giving rise to \eqref{E:main_result_recoverability_alpha_1} in Theorem~\ref{T:noise_free}.

\end{appendices}

\bibliographystyle{IEEEtran}
%
\bibliography{citations}

\end{document}